\documentclass[aps,prb,twocolumn,showpacs,superscriptaddress]{revtex4-2}  
\usepackage{graphicx}
\usepackage{amssymb}  
\usepackage{amsmath}  
\usepackage{amsfonts}
\usepackage{comment}
\usepackage{csquotes}
\usepackage{subfigure}
\usepackage[colorlinks,allcolors=blue]{hyperref}
\usepackage{braket}
\usepackage{xcolor}
\definecolor{added}{RGB}{0, 102, 204}  
\definecolor{deleted}{RGB}{204, 0, 0} 
\newcommand{\add}[1]{\textcolor{black}{#1}}

\begin{document}
\title{Fabry--Pérot interferometry with stochastic anyonic sources}

\author{Sarthak Girdhar}
\affiliation{Department of Physics, University of Basel, Klingelbergstrasse 82, 4056, Basel, Switzerland}

\author{Edvin G. Idrisov}
\affiliation{Department of Physics, United Arab Emirates University, P.O. Box 15551 Al-Ain, United Arab Emirates}

\author{Thomas L. Schmidt}
\affiliation{Department of Physics and Materials Science, University of Luxembourg, 1511 Luxembourg, Luxembourg}

\begin{abstract}
We investigate the interference of Laughlin quasiparticles (QPs) in the fractional quantum Hall regime that are stochastically injected into a Fabry--P\'erot interferometer. We find that the effective Aharonov--Bohm (AB) phase accumulated along the interferometer loop acquires an additional contribution of $\sin(2\pi\lambda)/2$ per QP present on it, where $\pi\lambda$ is the QP exchange phase. This contribution originates from time-domain braiding processes associated with injected QPs passing the interferometer quantum point contacts. In the limit of symmetric QP injection, the tunneling current noise exhibits AB oscillations as a function of the total injected current, providing access to the exchange phase $\pi\lambda$. In the regime of large total injection, we identify a universal Fano factor that displays power-law scaling and a characteristic phase shift reflecting \add{time-domain QP braiding at the interferometer QPCs}. These results are relevant for accessing anyonic exchange statistics in mesoscopic interferometers. 
\end{abstract}

\maketitle

\section{Introduction}
Fermions and bosons are the fundamental classes of particles distinguished by their exchange statistics. In three dimensions, no other kinds of exchange statistics are possible due to the topology of the underlying space~\cite{Leinaas1977}. However, in systems which are effectively two-dimensional, quasiparticles (QPs) with statistics between fermions and bosons can also emerge. These QPs are called anyons~\cite{Wilczek1982,Wilczek1982Anyon}. The canonical setup for realizing anyons is a fractional quantum Hall state \cite{Tsui}, where as a result of strong electron-electron interactions, they emerge as edge excitations with fractional charge and statistics~\cite{Saminadayar1997,dePicciotto1997,Carrega2021Anyons,Schiller2022Anyon,Goldman1995}. Recent advancements in hybrid mesoscopic devices have made it possible to study these excitations with unprecedented precision~\cite{HeiblumFeldman,Das2012,Dolev2008,Mitali1,Banerjee2018,Mitali}. These include the fractional quantum Hall states~\cite{HeiblumStern,Ezawa2013}, fractional quantum anomalous Hall states~\cite{Laughlin,Spanton2017Observation,Xu2023Observation,Nagaosa2009Anomalous} and fractional topological insulators~\cite{Peterson2021Trapped,Zhang2020Experimental,Breunig2021Opportunities,Maciejko2010}.

Along these lines, quantum point contacts (QPCs) have emerged as especially powerful tools, which let us control and probe electronic channels at the nanoscale~\cite{Kane1992,Ponomarenko2024Unusual,Liang,Kiczynski2022Engineering,Miller2007Fractional}. By tuning transmission and only selectively allowing interference between these different channels, they can reveal several fundamental quantum properties, from QP exchange statistics to phase coherence and conductance properties of sub-gap states. Although interpreting current-voltage characteristics from these devices is often challenging due to factors such as disorder and electron-phonon interactions~\cite{Kane1,Kane2,Mielke,Mross,Shytov1,Ashoori,Shytov2,IdrisovPhonon,Biswas2022}, QPCs can in principle be used to determine the fractional charge and exchange statistics of QPs. \add{This can be done through shot noise measurements, anyonic collisions, or
interferometry~\cite{Ezawa2013,Halperin2020}. On the theoretical side, interferometric probes of anyonic
statistics have been proposed in Fabry--P\'erot~\cite{two_point_contact_interferometer,Kivelson2025}, Mach--Zehnder~\cite{Law2006}, and Hanbury Brown--Twiss~\cite{Campagnano2012,Safi2001} geometries. Experimentally,
anyonic interference has by now been observed in Fabry--P\'erot
interferometers in GaAs~\cite{Ghosh2025Bunching,nakamura2023fabryperot} and graphene-based devices~\cite{Samuelson2026,Kim2024,Werkmeister2025},
as well as in Mach--Zehnder interferometers~\cite{Kundu2023,Ghosh2025MZ}.}

\begin{figure}
\includegraphics[width=\columnwidth]{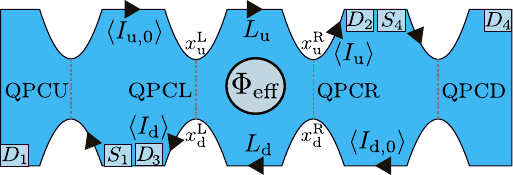}
\caption{
Schematic of an anyonic Fabry--P\'erot interferometer (FPI). The blue region denotes the fractional quantum Hall fluid. Dilute beams of Laughlin quasiparticles (QPs) are injected via the source contacts, QPCU and QPCD, from sources $S_1$ and $S_4$, with average currents $\langle I_{{\rm u},0}\rangle$ and $\langle I_{{\rm d},0}\rangle$, respectively. The QPs then backscatter and interfere via QPCL and QPCR, forming an interference loop of total perimeter $L_+ = L_{\rm u} + L_{\rm d}$, where $L_{\rm u,d}$ are the lengths of the loop arms. $\Phi_{\rm eff}$ represents the total effective Aharonov--Bohm phase accumulated upon traversing the loop once. In the nonequilibrium steady state, the outgoing currents $\langle I_{\rm u} \rangle$ and $\langle I_{\rm d} \rangle$ are detected at $D_2$ and $D_3$, respectively.
}\label{fig:FPI_HallBar_Schematic}
\end{figure}

Notably, recent experimental breakthroughs have enabled the construction of anyonic colliders at both integer and fractional filling factors~\cite{Gwendal2020}. Their measurements of zero-frequency current cross-correlations and the generalized Fano factor at fractional fillings have demonstrated possible evidence for anyonic exclusion statistics of charge carriers. 
These findings align with theoretical predictions for the Laughlin state at $\nu = 1/3$
~\cite{Rosenow}. An essential distinction in this experiment is the use of nonequilibrium source QPCs for individual QP injection, instead of conventional voltage sources~\cite{Ezawa2013,IdrisovAmmeter,IdrisovSchmidt} which often produce bunched QP flows where ambiguous low-frequency noise measurements obscure the distinction between fractional charges and electrons~\cite{Bisognin2019,Glidic2023}. 

Overall, the experimental realization of anyonic colliders and interferometers has sparked numerous investigations in both experimental and theoretical domains ~\cite{Taktak2022,Sim,Han,Mora1,Mora2,Lee2022,Lee2023,Feve2023,Bartolomei2022,Nakamura2020,Liang,nakamura2023fabryperot,Glidic2023,Glidic,Mitali,Kundu2023,Chamon2,PhysRevB.110.L041402,zhang2025effective,PhysRevLett.134.096303,Ji2003}. Motivated by this progress, in this work we study the interference of Laughlin QPs injected via nonequilibrium source contacts into a Fabry--P\'erot interferometer (FPI). Using the nonequilibrium bosonization technique, we calculate the tunneling current, its noise and the outgoing current cross-correlation. The stochastic nature of the injection process has several interesting effects on the interferometer. In particular, we show that stochastic injection introduces a current-dependent contribution to the effective Aharonov--Bohm (AB) phase arising from time-domain braiding at the interferometer QPCs. This enables pure AB oscillations in the current noise as a function of total injected current (without varying the magnetic field or the system geometry), with a periodicity directly related to the anyonic exchange phase. In the large-injection limit, an effective Fano factor emerges that encodes a characteristic phase shift from time-domain braiding of QPs around the interferometer QPCs, offering a robust and dimensionless signature of anyonic statistics.

This paper is organized as follows. In Sec.~\ref{Sec: Model_and_Summary}, we introduce our model and present a qualitative discussion of the main results, focusing on analogies between an optical and an anyonic FPI, as well as on AB oscillations in the latter. In Sec.~\ref{Sec: Formalism and Results}, we describe the model Hamiltonian, the nonequilibrium bosonization technique, and the correlation functions relevant for our setup. Building on this formalism, in Secs.~\ref{Sec:tunnelingcurrent} and~\ref{noise_and_cross_correlations} we compute the tunneling current, its noise, and the outgoing currents' cross-correlator in the nonequilibrium steady state. In Sec.~\ref{sec:AB_oscillations}, we analyze AB oscillations in the current noise and discuss signatures of anyonic exchange statistics through a Fano factor. We conclude with a summary and an outlook in Sec.~\ref{Sec:Conclusion}.

\section{Model and summary of the main results}
\label{Sec: Model_and_Summary}
 Our setup consists of an anyonic FPI in a Hall-bar geometry, as shown in Fig.~\ref{fig:FPI_HallBar_Schematic}. Dilute beams of Laughlin QPs are injected into the \enquote{up} $(\mathrm u)$ and \enquote{down} $(\mathrm d)$ arms from sources $S_1$ and $S_4$, respectively, via weak backscattering at the source contacts QPCU and QPCD. The corresponding average injected currents are respectively denoted by $\langle I_{{\rm u},0} \rangle$ and $\langle I_{{\rm d},0} \rangle$. The injected QPs subsequently propagate along the edges and interfere through tunneling at the interferometer QPCs labeled as QPCL and QPCR, forming an interference loop of total perimeter $L_+ = L_{\rm u} + L_{\rm d}$. Here, $L_{\rm u} = x^{\rm R}_{\rm u} - x^{\rm L}_{\rm u}$ and $L_{\rm d} = x^{\rm L}_{\rm d} - x^{\rm R}_{\rm d}$ denote the path lengths along the up and down arms, respectively.

All gates defining the QPCs are operated in the weak pinch-off regime, which is essential for QP tunneling. In the opposite limit of nearly opaque barriers, tunneling occurs only via electrons, precluding access to the exchange properties of fractional QPs~\cite{Kane_2003}.

As a QP propagates along the FPI loop, it acquires an AB phase, giving rise to quantum interference effects. In general, this phase depends on the applied magnetic field, the area enclosed by the interferometer loop, and the number of QPs localized within it. The latter can be controlled, for example, by a central gate that selectively depletes charge from the interior of the loop. As we demonstrate below, the total AB phase can also acquire additional contributions due to the injected currents $\langle I_{{\rm u},0} \rangle$ and $\langle I_{{\rm d},0} \rangle$. This provides a means to probe AB oscillations and anyonic exchange effects without varying the magnetic field or the geometric parameters of the interferometer.

In the nonequilibrium steady state, interference at QPCL and QPCR results in the tunneling of a net current $\langle I_{\rm T}\rangle$ from edge ${\rm u}$ to ${\rm d}$. The resulting outgoing currents $\langle I_{\rm u} \rangle$ and $\langle I_{\rm d} \rangle$ are measured at detectors $D_2$ and $D_3$, respectively.

\subsection{Coherent and incoherent quasiparticle interference}

In a classical optical FPI, the sharpest interference fringes are obtained with a monochromatic light source. For a finite spread $\Delta k$ around a central wavenumber $k$, the visibility of the interference fringes is controlled by the dimensionless product $d\,\Delta k$, where $d$ denotes the mirror separation. It is therefore instructive to analyze an anyonic FPI from a similar perspective.

We consider the injection of a QP into one of the interferometer arms through a source constriction, from where it propagates toward the tunneling QPCs QPCL and QPCR. In equilibrium, QP--quasihole (QP--QH) pairs are spontaneously generated at both QPCL and QPCR at all times, but owing to electron--hole symmetry, these processes give rise to a zero net current. In the presence of QP injection, however, the situation changes qualitatively. When an injected QP impinges on a QPC, a nontrivial interference arises between processes in which a QP--QH pair is created before and after the arrival of the injected QP at the QPC. This interference corresponds to a braiding process in the time domain \cite{Ruelle2025,Schiller2022Anyon,Rosenow2025,PhysRevLett.132.216601,Han2016,Sim,Mora1,Mora2}, and results in a statistical phase factor $e^{2\pi i\lambda}$, where $\pi\lambda$ is the QP exchange phase. 

For multiple injection events occurring randomly in time with an average injection rate $\langle I_{\alpha,0}\rangle/e^\ast$ ($\alpha\in\{\mathrm{u},\mathrm{d}\}$), where $e^\ast=e/m$ is the charge of a Laughlin QP, $e$ is the electron charge and $m$ is an odd number, the accumulated statistical phase fluctuates both temporally and spatially along the edge. Since the injection events follow a Poissonian distribution, these fluctuations are characterized by the average rate $\langle I_{\alpha,0}\rangle/e^\ast$. It is therefore convenient to describe the interferometer in terms of the dimensionless parameters
\begin{equation} \label{kappa_u_and_d_definition}
N_{\rm u}=\frac{L_+\langle I_{{\rm u},0}\rangle}{v e^\ast},\qquad
N_{\rm d}=\frac{L_+\langle I_{{\rm d},0}\rangle}{v e^\ast},
\end{equation}
where $v$ is the renormalized QP speed along both edges.

Let $\zeta_{\rm L}$ and $\zeta_{\rm R}$ denote the backscattering amplitudes at QPCL and QPCR, respectively. In the limits of both small and large $N_{\rm u}$ and $N_{\rm d}$, the two interferometer QPCs can therefore be effectively replaced by a single QPC with an effective amplitude $\zeta_{\rm eff}$ given by
\begin{equation}
|\zeta_{\rm eff}|^2=
\begin{cases}
|\zeta_{\rm L}+\zeta_{\rm R}|^2, & N_{\rm u/d}\ll 1,\\[4pt]
|\zeta_{\rm L}|^2+|\zeta_{\rm R}|^2, & N_{\rm u/d}\gg 1,
\end{cases}
\end{equation}
corresponding to the coherent and incoherent interference regimes, respectively. In the intermediate regime, $N_{\rm u/d}\sim\mathcal{O}(1)$, one expects damped AB oscillations. This damping arises from the combined effects of the characteristic power-law decay of edge state correlation functions and the dephasing induced by stochastic QP injection processes.

\subsection{Aharonov--Bohm oscillations and fractional exchange statistics}
As the injected QPs backscatter at QPCL and QPCR and traverse the FPI loop, they acquire an AB phase in the process. It is generally agreed that this phase has three components~\cite{Feldman2021Fractional,feldman_robustness_2022,PhysRevLett.98.106801,PhysRevB.83.155440}: a nonuniversal component from backscattering at the QPCs, a contribution determined by the magnetic flux through the loop, and a statistical phase which depends on the filling factor and the total number of QPs in the bulk of the loop. At a constant filling factor, magnetic field and gate voltages responsible for the QPCs, these contributions are essentially constant. It is therefore convenient to incorporate all of these contributions into the relative phase of the tunneling amplitudes at QPCL and QPCR, $\zeta_{l}$ with $l\in\{\mathrm{L},\mathrm{R}\}$, according to
\begin{equation}
    \zeta_{\rm L}^\ast \zeta_{\rm R}
    =|\zeta_{\rm L}||\zeta_{\rm R}|e^{ i\Phi}.
\end{equation}
However, in addition to these equilibrium contributions, QP injection at the source contacts also introduces an additional phase component. As discussed above, individual injection events generate random phase fluctuations along the edge. In the nonequilibrium steady state, these fluctuations give rise to an average phase per QP residing on the interferometer arms. Using the nonequilibrium bosonization technique, which is nonperturbative in the injection QPCs, we find this contribution to be $\sin(2\pi\lambda)/2$, leading to an effective phase
\begin{equation} \label{phi_effective}
    \Phi_{\rm eff}=\Phi+\frac{N}{2}\sin(2\pi\lambda),
\end{equation} where \begin{equation}N=N_{\rm u}+N_{\rm d}\end{equation} corresponds to the average total number of injected QPs on the interferometer arms.

We denote the interference contributions to the tunneling current, its zero-frequency noise and the outgoing currents' cross-correlator as $\langle I_{\rm int}\rangle$, $\langle S_{\rm int}\rangle$ and $\langle K_{\rm int}\rangle$, respectively. Under the symmetric biasing condition, i.e., $I_-=\langle I_{{\rm u},0} \rangle-\langle I_{{\rm d},0}
\rangle=0$, we find that $\langle S_{\rm int}\rangle$ exhibits pure AB oscillations as $N$ is varied:
\begin{equation} \label{eq:AB_oscillations_summary}
    \langle S_{\rm int} \rangle\big|_{I_-=0} \propto \cos(\Phi_{\rm eff}).
\end{equation}
Importantly, $N$ can be tuned by adjusting the total injected current $I_+=\langle I_{{\rm u},0}\rangle+\langle I_{{\rm d},0}\rangle$, while keeping $L_+$, $e^\ast$, and hence $\Phi$ fixed. 

\add{Additionally, time-domain braiding of QP--QH pairs created at the two
interferometer QPCs [see Sec.~\ref{sec:AB_oscillations} B] gives rise to a characteristic phase
shift of $\pi\lambda$ on top of the background AB phase $\Phi_{\rm eff}$.
We find that this phase shift can be observed through an effective Fano
factor $P$, defined as}~\cite{Rosenow}
\begin{equation}
    P\!\left(\frac{I_-}{I_+},N\right)
    =
    \frac{
        \left\langle K_{\rm int} \right\rangle
    }{
        e^\ast I_+
        \left.
        \dfrac{\partial}{\partial I_-}
        \langle I_{\rm int} \rangle
        \right|_{I_-=0}
    }.
    \label{eq:fano_factor_definition_summary}
\end{equation} In the limit $I_-=0$ and $N \to \infty$, this ratio obeys
\begin{equation}
\label{fano_factor_phase_shift_summary}
   1- P\propto
    \frac{
        \cos\!\left(\Phi_{\rm eff}\right)
    }{
        \cos\!\left(\Phi_{\rm eff}+\pi\lambda\right)
    }.
\end{equation} Equations~\eqref{eq:AB_oscillations_summary} and~\eqref{fano_factor_phase_shift_summary} constitute the two central results of this work and, in principle, allow for a direct measurement of the anyonic exchange phase $\pi\lambda$ in the proposed setup.

\section{Formalism and approach} \label{Sec: Formalism and Results}
\subsection{Bosonization} \label{Subsec:Bosonization}
In order to describe the one-dimensional edges $\rm u$ and $\rm d$, we use the bosonization technique \cite{Giamarchi2004}. For now we employ equilibrium bosonization; the effects of the injection QPCs will be considered in the next subsection. The anyonic field operator $\psi_\alpha(x)$ on an edge is \begin{equation} \label{definition of quasiparticle field operator}
\psi_\alpha(x)=\frac{1}{\sqrt{2\pi a}}e^{i\phi_\alpha(x)}, \
    \alpha\in\{\mathrm{u},\mathrm{d}\},
\end{equation} 
where $a$ is a short-distance ultraviolet cutoff. The corresponding QP charge density is
\begin{equation} \label{eq:charge_density}
    \rho_\alpha(x)=\frac{e}{2\pi}\partial_{x}\phi_\alpha(x).
\end{equation} The field $\phi_\alpha(x)$ is a chiral bosonic field satisfying the equal-time commutation relation
\begin{equation}
\label{bosonic_equal_time_commutator}
[\phi_\alpha(x_1),\phi_\beta(x_2)]
    = i\pi\lambda\delta_{\alpha\beta}\,\operatorname{sign}(x_1-x_2),
\end{equation} where $\lambda=1/m$ for a Laughlin state. In general, edge reconstruction or impurity-induced dephasing may result in $e\lambda \neq e^\ast$ \cite{Halperin2002}. However, in the present work we do not consider these complications. We note that no chirality index appears in the commutation relation in Eq.~\eqref{bosonic_equal_time_commutator}, since the coordinate systems on both edges are defined to increase in the direction of QP propagation. As a result, all modes are effectively right-moving, and a chirality label is unnecessary. 

The effective Hamiltonian of the system is written as
\begin{equation}
    H=\sum_{\alpha} H_\alpha+H_{\rm T},
\end{equation}
where the first term describes free, chiral Laughlin QP modes on edge $\alpha$ and is given by
\begin{equation}
    H_\alpha
    =\frac{v}{4\pi\lambda}\int dx\,\bigl[\partial_x \phi_\alpha(x)\bigr]^2.
\end{equation} The second term, $H_{\rm T}$, describes weak QP tunneling between the upper and lower edges through QPCL and QPCR, and takes the form
\begin{equation}
\label{tunnelingHamiltonian}
    H_{\rm T}=\sum_{l\in\{\mathrm{L},\mathrm{R}\}}\left(A_l+A_l^\dagger\right),
\end{equation}
where $A_l$ denotes the operator that transfers a fractional charge $e^\ast$ from the upper to the lower edge through the $l^{\rm th}$ QPC. Explicitly,
\begin{equation}
\label{Tunneling vertex operator}
    A_l
=\zeta_{l}\,e^{i\phi_{\rm u}(x_{\rm u}^l)-i\phi_{\rm d}(x_{\rm d}^l)}.
\end{equation}
It is worth emphasizing that the backscattering amplitudes $\zeta_{l}$ are complex numbers rather than operators. Strictly speaking, locality of the tunneling operators associated with spatially separated QPCs would normally require the introduction of Klein factors in the definition of the QP field operators in Eq.~\eqref{definition of quasiparticle field operator}, ensuring that $[A_{\rm L},A_{\rm R}]=0$. However, using the equal-time commutation relation in Eq.~\eqref{bosonic_equal_time_commutator}, one finds that in the absence of Klein factors,
\begin{equation}
\label{AL_AR_swap}
    A_{\rm L} A_{\rm R}
=e^{i\pi\lambda\left[\operatorname{sign}(x_{\rm u}^{\rm R}-x_{\rm u}^{\rm L})
    +\operatorname{sign}(x_{\rm d}^{\rm R}-x_{\rm d}^{\rm L})\right]}A_{\rm R} A_{\rm L}.
\end{equation}
For a Fabry--P\'erot geometry, the sign functions in the exponent of Eq.~\eqref{AL_AR_swap} cancel exactly~\cite{MZI_interferometer,PhysRevB.86.245105}, see Fig.~\ref{fig:FPI_HallBar_Schematic}, implying that $A_{\rm L}$ and $A_{\rm R}$ commute. The tunneling Hamiltonian in Eq.~\eqref{tunnelingHamiltonian} therefore already satisfies the required locality conditions, and we proceed without introducing Klein factors.

We now go to the interaction picture with $H_{\rm T}$ acting as a perturbation. The bosonic fields evolve as 
\begin{equation} \label{eq:boson_dynamics}
\phi_\alpha(x,t)=\phi_\alpha(x-vt,0).
\end{equation}  
Using Eq.~\eqref{bosonic_equal_time_commutator}, we can then write down the time-dependent commutation relation
\begin{equation} \label{bosoniccommutationrelation_in_time}[\phi_\alpha(x_1,t_1),\phi_\beta(x_2,t_2)]=i\pi\lambda\delta_{\alpha\beta}\ \text{sign}\left[x_1-x_2-v(t_1-t_2)\right].
\end{equation} 
When the difference $I_-=\langle I_{{\rm u},0}\rangle-\langle I_{{\rm d},0}\rangle$
is nonzero, the presence of QPCL and QPCR allows a net QP current to tunnel between the edges $\mathrm{u}$ and $\mathrm{d}$. This tunneling current is defined as
\begin{equation} \label{eq:current_op_defintion}
    I_{\rm T}
    = ie^\ast[H_{\rm T},n_{\rm d}]
    = i e^\ast \sum_{l}\left(A_l^\dagger-A_l\right),
\end{equation}
where
\add{\begin{equation}
    n_{\rm d}(t)=\int \frac{dx}{2\pi\lambda} \partial_x\phi_{\rm d}(x,t)
\end{equation}}
is the total number of QPs on the lower edge. It satisfies the current conservation equation,
\begin{equation} \label{eq:tunnel_current_definition}
    \langle I_{\rm T}\rangle
    =\langle I_{{\rm u},0}\rangle-\langle I_{\rm u}\rangle
    =\langle I_{\rm d}\rangle-\langle I_{{\rm d},0}\rangle.
\end{equation}

\subsection{Nonequilibrium correlation functions}  The equilibrium correlation function of the tunneling operators is given by~\cite{Giamarchi2004}
\begin{align}
\label{eq:equilibrium_correlation}
\langle A_l(t)\, A_{l'}^{\dagger}(t') \rangle_{\mathrm{eq.}}
&= \zeta_{l} \zeta_{l'}^{*}\,
   \tau_c^{2\lambda}\,
\frac{e^{-\frac{i \pi \lambda}{2} \text{sign}
    \left(t - t' - (x^l_{\rm u} - x^{l'}_{\rm u}) / v\right)}}{\bigl|t - t' - (x_{\rm u}^{l} - x_{\rm u}^{l'})/v - i\tau_c \bigr|^{\lambda}}
   \nonumber \\[4pt]
&\quad \times
\frac{e^{-\frac{i \pi \lambda}{2} \text{sign}
    \left(t - t' - (x^l_{\rm d} - x^{l'}_{\rm d})/ v\right)}}{\bigl|t - t' - (x_{\rm d}^{l} - x_{\rm d}^{l'})/v + i\tau_c \bigr|^{\lambda}},
\end{align}
where $(l,l')\in\{\rm L,R\}$ and $\tau_c \propto a/v$ is a short-time cutoff. The average is taken with respect to the equilibrium ground state at zero temperature~\footnote{Strictly speaking, the temperature is sufficiently low for the above result to be valid, but still satisfies $|\zeta_{\rm L,R}|\ll T^{1-\lambda}$ such that a perturbative treatment of the interferometer QPCs remains valid, see Ref.~\cite{Kane_2003}.}. 

We now incorporate the effect of the source QPCs on these correlation functions. A naive approach based on perturbation theory in the corresponding backscattering amplitudes fails, as such an expansion is known to be divergent \cite{noise_induced_PT,Ivan_Noneq_Bosonization,Kane_2003}. A nonperturbative description of the injection process is therefore required~\cite{Rosenow,GutmanZero,Gutman}.

Consider rare, instantaneous, and Markovian injection of QPs into edge $\alpha$ through a source QPC located at position $x_{\rm inj}$. For a single injection event occurring at time $t_{\rm inj}$, the QP charge density undergoes the shift 
\begin{equation}
    \rho_\alpha(x,t_{\rm inj})
    \to
    \rho_\alpha(x,t_{\rm inj})+e^\ast\, \delta(x-x_{\rm inj}),
\end{equation} 
where $\delta(x-x_{\rm inj})$ denotes the Dirac delta function.
Using Eq.~\eqref{eq:charge_density}, this corresponds to a kink in the bosonic field,
\begin{equation} \label{bosonic_field_shift}
    \phi_\alpha(x,t_{\rm inj})
    \to
    \phi_\alpha(x,t_{\rm inj})+2\pi\lambda\,\Theta(x-x_{\rm inj}).
\end{equation} For chiral propagation, the corresponding field configuration at arbitrary time $t$ takes the form
\begin{equation}
    \phi_\alpha(x,t)
    \to
    \phi_\alpha(x,t)+2\pi\lambda\,\Theta\ \!\bigl(x-x_{\rm inj}-v(t-t_{\rm inj})\bigr).
\end{equation}
Thus, the effect of the source QPCs can be treated nonperturbatively by representing each injected QP as a soliton propagating along the edge. This c-number shift manifests itself in correlation functions as a phase factor.

In particular, the equilibrium correlator in Eq.~\eqref{eq:equilibrium_correlation} is modified according to
\begin{align}
\label{eq:correlation_phase_shift}
    \langle e^{i\phi_\alpha(x,t)} e^{-i\phi_\alpha(x,t')} \rangle
    \to\;
    &\langle e^{i\phi_\alpha(x,t)} e^{-i\phi_\alpha(x,t')} \rangle \nonumber \\
    &\times e^{-2\pi i\lambda\,[\Theta(t-t_0)-\Theta(t'-t_0)]},
\end{align}
where
\begin{equation}
    t_0=t_{\rm inj}+\frac{x-x_{\rm inj}}{v}
\end{equation}
is the arrival time of the injected QP at position $x$. The correlator thus acquires a phase factor $e^{- 2\pi i \lambda \text{sign}(t-t')}$ if and only if the injected QP crosses the point $x$ between times $t$ and $t'$, and remains unchanged otherwise. This phase accumulation is the essence of time-domain braiding~\cite{Schiller2022Anyon}: the arrival of a QP between $t'$ and $t$ corresponds to a closed loop in time enclosing a statistical flux $2\pi\lambda$.

For multiple injection events occurring independently and following Poissonian statistics, with average rates $\langle I_{{\rm u},0}\rangle/e^\ast$ and $\langle I_{{\rm d},0}\rangle/e^\ast$ on the upper and lower edges, respectively, one performs a statistical average. This yields
\begin{equation}
\label{nonequilibriumcorrelationfunction_l=l'}
    \langle A_l(t)A_l^\dagger(0)\rangle_0
    =
    \langle A_l(t)A_l^\dagger(0)\rangle_{\rm eq.}
    \,\chi_{\rm u}(t)\chi_{\rm d}(t),
\end{equation}
where the subscript $0$ denotes an out-of-equilibrium scenario and $\chi_\alpha(t)$ are the generators of full counting statistics \cite{Kambly2010Factorial,Landi2023Current,PhysRevLett.105.256802,PhysRevB.87.195433,PhysRevB.91.245419}. Explicitly,
\begin{equation}
\label{counting_statistics_generator_u}
    \chi_{\rm u}(t)
    =
    \exp\!\left[
    -\frac{\langle I_{{\rm u},0}\rangle |t|}{e^\ast}
    \bigl(1-e^{-2\pi i\lambda\,\operatorname{sign}(t)}\bigr)
    \right],
\end{equation}
and
\begin{equation}
\label{counting_statistics_generator_d}
    \chi_{\rm d}(t)
    =
    \exp\!\left[
    -\frac{\langle I_{{\rm d},0}\rangle |t|}{e^\ast}
    \bigl(1-e^{2\pi i\lambda\,\operatorname{sign}(t)}\bigr)
    \right],
\end{equation} where we assumed $\langle I_{\rm u,0}\rangle,\langle I_{\rm d,0}\rangle \geq 0$. Similarly, the nonlocal correlators take the form
\begin{align}
\label{nonequilibrium_rl_correlations}
    \langle A_{\rm R}(t)A_{\rm L}^\dagger(0)\rangle_0
    &=
    \langle A_{\rm R}(t)A_{\rm L}^\dagger(0)\rangle_{\rm eq.} \notag \\
    &\times \chi_{\rm d}\!\left(t+\frac{L_{\rm d}}{v}\right)
    \chi_{\rm u}\!\left(t-\frac{L_{\rm u}}{v}\right).
\end{align} 

Injection events thus introduce stochastic phase shifts in the edge. For Poissonian injection, the variance of these phase fluctuations scales with the mean injection rate, implying that the real part of $\log\chi_\alpha(t)$ leads to exponential dephasing at large currents and long times. The imaginary part, as we show below, gives rise to an effective contribution to the AB phase.

\add{It is interesting to note that for integer filling factor $\lambda$, the nonequilibrium shift in Eq.~\eqref{eq:correlation_phase_shift} vanishes, and the generators in Eqs.~\eqref{counting_statistics_generator_u} and \eqref{counting_statistics_generator_d} become independent of $\langle I_{\alpha,0}\rangle$ and reduce to
unity. Due to the resultant symmetry between the lower and the upper FPI arms, this means that the tunneling current vanishes identically~\cite{PhysRevLett.130.186203}. This reflects the fact that, within our idealized model of point-like injection and point-like tunneling, injected
charges affect the QP--QH pair creation at an interferometer QPC only through the
time-domain braiding phase $e^{2\pi i\lambda}$, which becomes trivial for
electrons. A nonvanishing tunneling current in the case of free fermions would require a finite
injection pulse width, finite contact size, or charging effects, which are beyond
the scope of this work~\cite{PhysRevLett.132.216601,PhysRevLett.130.186203,PhysRevLett.132.156501}. Our results therefore apply to fractional $\lambda = 1/m$ for odd integers $m \geq 3$.}

\section{Tunneling current}
\label{Sec:tunnelingcurrent} 
Using the definition of the tunneling current operator in Eq.~\eqref{eq:current_op_defintion}, the expectation value of the tunneling current to leading order in $H_{\rm T}$ is obtained from the Kubo formula,
\begin{equation}
\label{currentexpectationvalue}
    \langle I_{\rm T} \rangle
    = e^\ast \sum_{l,l'}
    \int dt\,
    \bigl\langle [A_l^\dagger(0),A_{l'}(t)] \bigr\rangle_0
    \equiv \sum_{l,l'} \langle I_{ll'} \rangle,
\end{equation}
where $[\,,\,]$ denotes a commutator. It is convenient to decompose this current as 
\begin{align}
\langle I_{\rm T} \rangle
    &= \langle I_{\rm dir} \rangle + \langle I_{\rm int} \rangle, \notag \\
    \langle I_{\rm dir} \rangle
    &= \langle I_{\rm LL} \rangle + \langle I_{\rm RR} \rangle, \notag \\
    \langle I_{\rm int} \rangle
    &= \langle I_{\rm LR} \rangle + \langle I_{\rm RL} \rangle,
\end{align}
where the direct contribution $\langle I_{\rm dir} \rangle$ arises from tunneling at the individual QPCs, while the interference contribution $\langle I_{\rm int} \rangle$ encodes the effects of tunneling processes involving both QPCL and QPCR.

We parametrize the injected currents on the upper and lower edges in terms of
\begin{equation}
    I_\pm = \langle I_{{\rm u},0} \rangle \pm \langle I_{{\rm d},0} \rangle.
\end{equation}
The direct tunneling current is then given by (refer to App.~\ref{App:Iint and Sint appendix} for details)
\begin{align}
\label{Result for direct tunneling current}
    \langle I_{\rm dir} \rangle
    &=
    C \bigl(|\zeta_{\rm L}|^2 + |\zeta_{\rm R}|^2\bigr)
    \Gamma(1-2\lambda)
    \left(\Omega^2+\xi^2\right)^{\lambda-\frac{1}{2}} \notag \\
    &\times
    \sin(\pi\lambda)\sin\!\left[(1-2\lambda)
    \tan^{-1}\!\left(\frac{\xi}{\Omega}\right)\right],
\end{align}
where $C= 4 \tau_c^{2\lambda} e^\ast$ and
\begin{align}
\label{eq:omega_xi_def}
    \Omega &= \frac{I_+}{e^\ast}\bigl[1-\cos(2\pi\lambda)\bigr], \notag \\
    \xi &= \frac{I_-}{e^\ast}\sin(2\pi\lambda).
\end{align}
The interference contribution takes the form (note that $\langle I_{\rm RL} \rangle=\langle I_{\rm LR} \rangle^*$)
\add{\begin{align}
\label{eq:interference_current_main}
    \langle I_{\rm int} \rangle
    &=
    C \left(\frac{L_+}{2v}\right)^{1-2\lambda}
    |\zeta_{\rm L}||\zeta_{\rm R}|\sin(\pi\lambda) \\
&\times 
    \operatorname{Im}\Bigl\{
    e^{i\Phi_{\rm eff}} 
    \Bigl[e^{-\mu} f(z^\ast)-e^{\mu} f(z)
    \Bigr]
    \Bigr\}, \notag 
\end{align}}
where we introduced 
\begin{equation} \label{f_z_definition_main_text}
    f(z)
    = 2^{\frac{1}{2}-\lambda}\pi^{-\frac{1}{2}}
    \Gamma(1-\lambda)\,
    z^{\lambda-\frac{1}{2}} K_{\frac{1}{2}-\lambda}(z),
\end{equation}
and $K_{\frac{1}{2}-\lambda}(z)$ denotes the modified Bessel function of the second kind. Moreover, we used
\begin{equation} \label{mu_and_z_definition}
    \mu=\frac{I_-L_+}{2ve^\ast}[1-\cos(2\pi\lambda)],\ z = \frac{L_+}{2v}\left(\Omega + i \xi \right).
\end{equation}
In both currents, we dropped correction terms of order $\tau_c^{2\lambda+1}$, since they vanish in the limit $\tau_c \to 0$, as expected from the RG irrelevance of the backscattering operators. \add{We note that as a function of $z$, the current $\langle I_{\rm int} \rangle$ is analytic in the open right half of the complex plane.} Since $\text{Re}(z) \geq 0$, the fractional powers of $z$ appearing in its definition are free of branch ambiguities. Finally, we note from Eq.~(\ref{eq:interference_current_main}) that as a result of the injection process, the effective phase around the FPI loop is given by
\begin{equation} \label{eq:phi_effective_main_text_definition}
    \Phi_{\rm eff}=\Phi+\frac{N}{2}\sin(2\pi\lambda).
\end{equation}
This expression can be interpreted in the following way: due to the Poissonian injection process, new QPs enter the FPI loop at time intervals of approximately $e^\ast/I_+$, such that $ve^\ast/I_+$ can be interpreted as the average distance between them. The ratio $N=I_+L_+/(ve^\ast)$ then corresponds to the average number of QPs on the loop, each contributing an effective phase $\sin(2\pi\lambda)/2$ due to time-domain braiding processes as it backscatters through QPCL and QPCR. By changing the total current injected into the loop, one can thus study pure AB oscillations in the setup. We will elaborate on this in Sec.~\ref{zero_bias_noise}.

\begin{figure}
    \centering
    \includegraphics[width=0.95\linewidth]{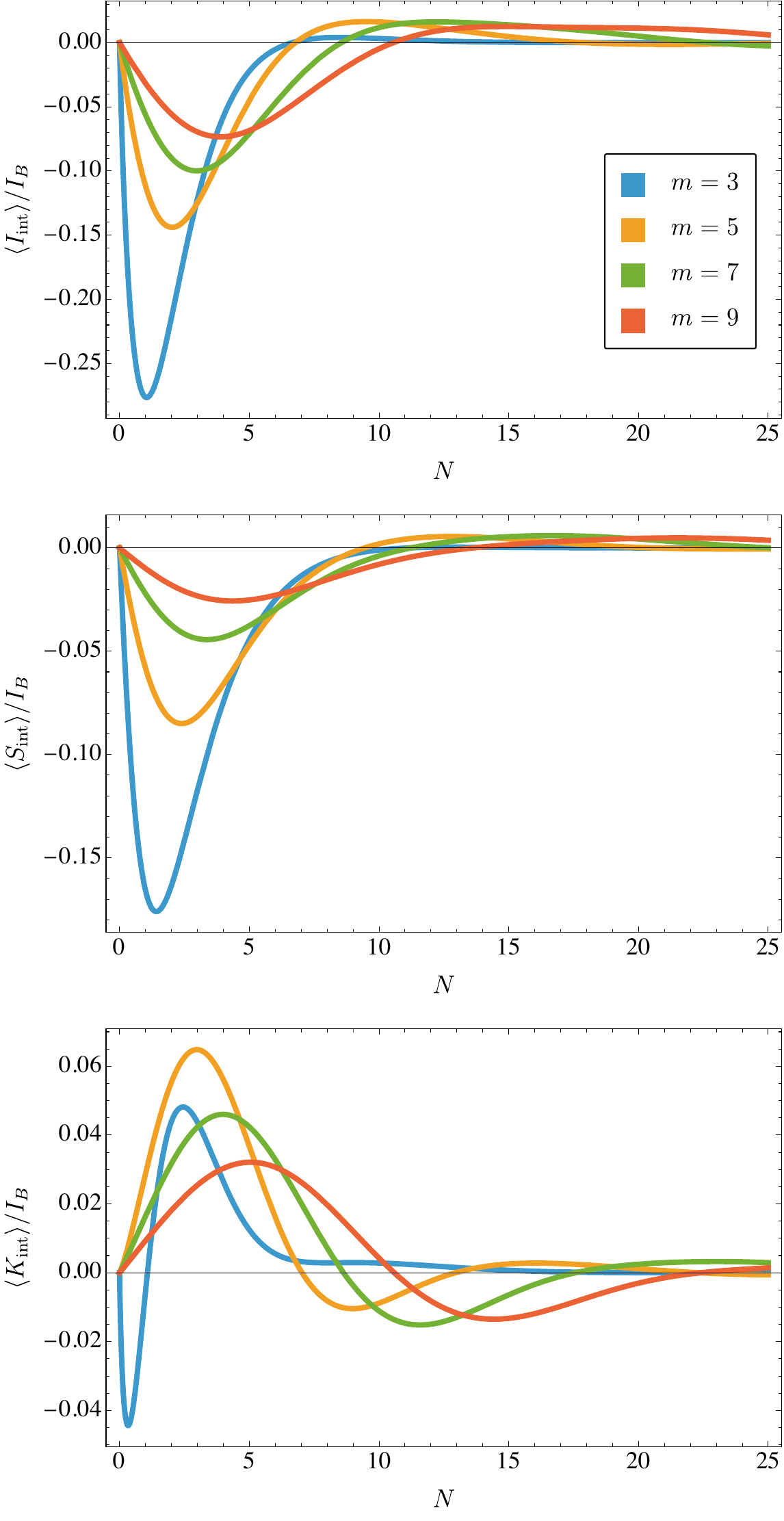}
\caption{Interference contributions $\langle I_{\rm int}\rangle$, $\langle S_{\rm int}\rangle$, and $\langle K_{\rm int}\rangle$ to the average tunneling current, its noise, and the outgoing current cross-correlator, shown as functions of $N = I_+ L_+ /(v e^\ast)$ for Laughlin filling fractions $e\lambda = e^\ast = e/m$. Here, $I_B = |\zeta_{\rm L}|\,|\zeta_{\rm R}|\,\tau_c^{2\lambda} I_+^{2\lambda-1}$, $I_- = 0.4 I_+$ and the bare AB phase $\Phi = \pi/2$. In the large-$N$ limit, interference effects are washed out due to the power-law decay of equilibrium correlation functions and random phase fluctuations  along the edges $\mathrm{u}$ and $\mathrm{d}$. The quantity $I_-$ captures a phase imbalance between the edges $\mathrm{u}$ and $\mathrm{d}$, such that increasing it from zero results in stronger dephasing compared to $I_-=0$, see Fig.~\ref{fig:Sint_AB_oscillations}.}
    \label{fig:plot_grid}
\end{figure}

\section{Current noise and cross-correlations}
\label{noise_and_cross_correlations}
We now turn to the fluctuations of the tunneling current. The zero-frequency noise is given by
\begin{equation}
\label{Expression for tunneling currents noise through anticommutator}
\langle \delta I_{\rm T}^2 \rangle_{\omega=0}
=(e^\ast)^2 \sum_{l,l^\prime}
\int dt\,
\langle \{A_l^\dagger(0),A_{l^\prime}(t)\} \rangle_0
=\sum_{l,l^\prime}\langle S_{ll^\prime} \rangle,
\end{equation}
where $\{\, , \,\}$ denotes an anticommutator. As before, we decompose the noise into a direct contribution
$\langle S_{\rm dir} \rangle=\langle S_{\rm LL}\rangle+\langle S_{\rm RR}\rangle$
and an interference contribution
$\langle S_{\rm int} \rangle=\langle S_{\rm LR}\rangle+\langle S_{\rm RL}\rangle$. The direct noise is found to be
\begin{equation}
\label{Final result for direct noise of tunneling current}
\begin{split}
\langle S_{\rm dir} \rangle
=&\ C'
\left(|\zeta_{\rm L}|^2+|\zeta_{\rm R}|^2\right)
\Gamma(1-2\lambda)
\left(\Omega^2+\xi^2\right)^{\lambda-\frac{1}{2}} \\
&\times
\cos(\pi\lambda)\cos\!\left[(1-2\lambda)\tan^{-1}\!\left(\frac{\xi}{\Omega}\right)\right],
\end{split}
\end{equation}
where $C' = 4 (e^\ast)^2 \tau_{\rm c}^{2\lambda}$. The interference contribution reads
\begin{align} \label{eq:interference_noise} 
    \langle S_{\mathrm{int}} \rangle 
&= 
    C' \left( \frac{L_+}{2v} \right)^{1-2\lambda}|\zeta_{\rm L}|\,|\zeta_{\rm R}| \nonumber
    \operatorname{Re} \Bigl\{ e^{i\Phi_{\mathrm{eff}}} \Bigl[ \cos(\pi\lambda) \\ 
&\times \left(e^{\mu} f(z) + e^{-\mu} f(z^\ast)\right) + e^{-\frac{L_+\Omega}{2v}} W(\tilde z)\Bigr]\Bigr\} 
, \end{align}
where 
\add{\begin{equation} \label{tilde_z_definition}
    \tilde z = \frac{L_+}{2v}\left(\frac{I_-}{I_+}\Omega - i \xi \right)
\end{equation}} and 
\begin{align} \label{eq: W(z)_definition}
    W(\tilde z) 
&=
    \int_{-1}^{1}du\ \frac{e^{u \tilde z}}{(1-u^2)^\lambda} \\
&=
    \sum_{k=0}^{\infty} \left(\frac{\tilde z}{2}\right)^{2k} \label{eq: W(z)_main_text}
    \frac{\sqrt{\pi}\,\Gamma(1-\lambda)} {k!\,\Gamma\!\left(k+\frac{3}{2}-\lambda\right)}.
\end{align}
We now turn to the zero-frequency cross-correlator of the outgoing currents,
\begin{equation}
\langle \delta I_{\rm u} \delta I_{\rm d} \rangle_{\omega=0}
= \int dt\, \langle \delta I_{\rm u}(t)\,\delta I_{\rm d}(0) \rangle .
\end{equation}
Using Eq. \eqref{eq:tunnel_current_definition}, the cross-correlator becomes
\begin{equation}
\langle \delta I_{\rm u} \delta I_{\rm d} \rangle_{\omega=0}
= - \langle \delta I_{\rm T}^2 \rangle_{\omega=0}
+ \langle \delta I_{{\rm u},0}\,\delta I_{\rm T} \rangle
- \langle \delta I_{\rm T} \, \delta I_{{\rm d},0}\rangle .
\end{equation}

For a Poissonian injection process at the source QPCs, this expression assumes the well-known form~\cite{Rosenow}
\begin{align}
\label{eq:cross_correlation_equation} \nonumber
\langle \delta I_{\rm u} & \delta I_{\rm d} \rangle_{\omega=0}
= - \langle \delta I_{\rm T}^2 \rangle_{\omega=0}
  \\ 
 &+ e^\ast
\Biggl[
\langle I_{{\rm u},0} \rangle
\frac{\partial}{\partial \langle I_{{\rm u},0} \rangle}
- \langle I_{{\rm d},0} \rangle
\frac{\partial}{\partial \langle I_{{\rm d},0} \rangle}
\Biggr]
\langle I_{\rm T} \rangle . 
\end{align} Equation~\eqref{eq:cross_correlation_equation} constitutes a nonequilibrium form of a fluctuation--dissipation relation. The first term captures noise generated at the interferometer QPCs (QPCL and QPCR), while the second term accounts for fluctuations originating from the injected currents $\langle I_{\alpha,0}\rangle$ at the source contacts QPCU and QPCD. As in the case of the tunneling current and its noise, the cross-correlator naturally decomposes into direct and interference contributions. Denoting its interference contribution by $\langle K_{\rm int}\rangle$, we find
\begin{align}
\label{eq:cross_correlation_equation_int} \nonumber
\langle K_{\rm int} \rangle
&= - \langle S_{\rm int} \rangle
  \\
 &+ e^\ast
\Biggl[
\langle I_{{\rm u},0} \rangle
\frac{\partial}{\partial \langle I_{{\rm u},0} \rangle}
- \langle I_{{\rm d},0} \rangle
\frac{\partial}{\partial \langle I_{{\rm d},0} \rangle}
\Biggr]
\langle I_{\rm int} \rangle . 
\end{align}
The explicit expression for $\langle K_{\rm int}\rangle$, as well as a derivation of Eq.~\eqref{eq:cross_correlation_equation} for our setup, are presented in Apps.~\ref{Appendix:interference terms in cross-correlations} and~\ref{Appendix:cross_correlation_derivation}, respectively. In the limit $L_+ \to 0$, we get fully coherent interference. Our results then reduce to those in Ref.~\cite{Rosenow}. We remark that although the calculations above are carried out explicitly for an interferometer with two QPCs, the resulting expressions for the tunneling current and its noise [see Eqs.~\eqref{currentexpectationvalue} and~\eqref{Expression for tunneling currents noise through anticommutator}]  take the form of sums over pairs of interferometer QPCs. This form allows for a straightforward generalization to an arbitrary number $n$ of interferometer QPCs, which we present in App.~\ref{n_QPCs}.

We show the dependence of $\langle I_{\rm int}\rangle$, $\langle S_{\rm int}\rangle$, and $\langle K_{\rm int}\rangle$ on $N$ in Fig.~\ref{fig:plot_grid}, all of which exhibit the expected damped oscillatory behavior. As discussed above, this damping originates from the combined effect of the power-law decay of Luttinger-liquid correlation functions and an additional exponential suppression due to QP injection at the source contacts. The oscillations arise from both the effective AB phase $\Phi_{\rm eff}$ and the imaginary part of the complex parameter $z$, see Eq.~\eqref{mu_and_z_definition}. The latter captures a phase imbalance between QPs injected into the edges $\mathrm{u}$ and $\mathrm{d}$, such that increasing the effective bias $I_-$ from zero enhances the randomness of phase fluctuations accumulated along the FPI arms, resulting in stronger dephasing. Consequently, the oscillations in Fig.~\ref{fig:plot_grid} are not purely AB in origin, but instead encode information about both QP statistics and transport properties. To isolate purely AB oscillations, we therefore set $I_- = 0$, as discussed in the following section.

\section{Aharonov--Bohm oscillations and fractional statistics} 
\label{sec:AB_oscillations}
\subsection{Current noise at zero effective bias: \texorpdfstring{$I_-=0$}{}}
\label{zero_bias_noise}
In the limit $I_-=0$, the tunneling current is obviously zero. However, its noise is
\begin{align}
\label{eq:S_int_Iminus0}
 \langle S_{\mathrm{int}} \rangle\big|_{I_- = 0}
&=    C' \left( \frac{L_+}{2v} \right)^{1-2\lambda} |\zeta_{\rm L}||\zeta_{\rm R}| \biggl[2 
   \cos(\pi\lambda) f\left(\frac{L_+\Omega}{2v}\right) 
   \nonumber  \\
& +
   W(0)e^{-\frac{L_+\Omega}{2v}} \biggr]\cos\left(\Phi_{\rm eff}\right).
\end{align} 
Equation~\eqref{eq:S_int_Iminus0} is one of the two main results of this paper. The zero effective bias current noise corresponds to AB oscillations as a function of $N$ with a constant frequency $\sin(2\pi\lambda)/2$. Therefore, even before the exponential dephasing suppresses all interference effects at $N \gg 1$, they can also be suppressed at the zeroes of the above noise, separated by half-period $2\pi / \sin(2\pi \lambda)$. If an experiment is designed to detect the components of the outgoing currents that change with $\Phi$, the above equation can in principle be used to measure $\pi\lambda$. 
\begin{figure}
    \centering
   \includegraphics[width=1\linewidth]{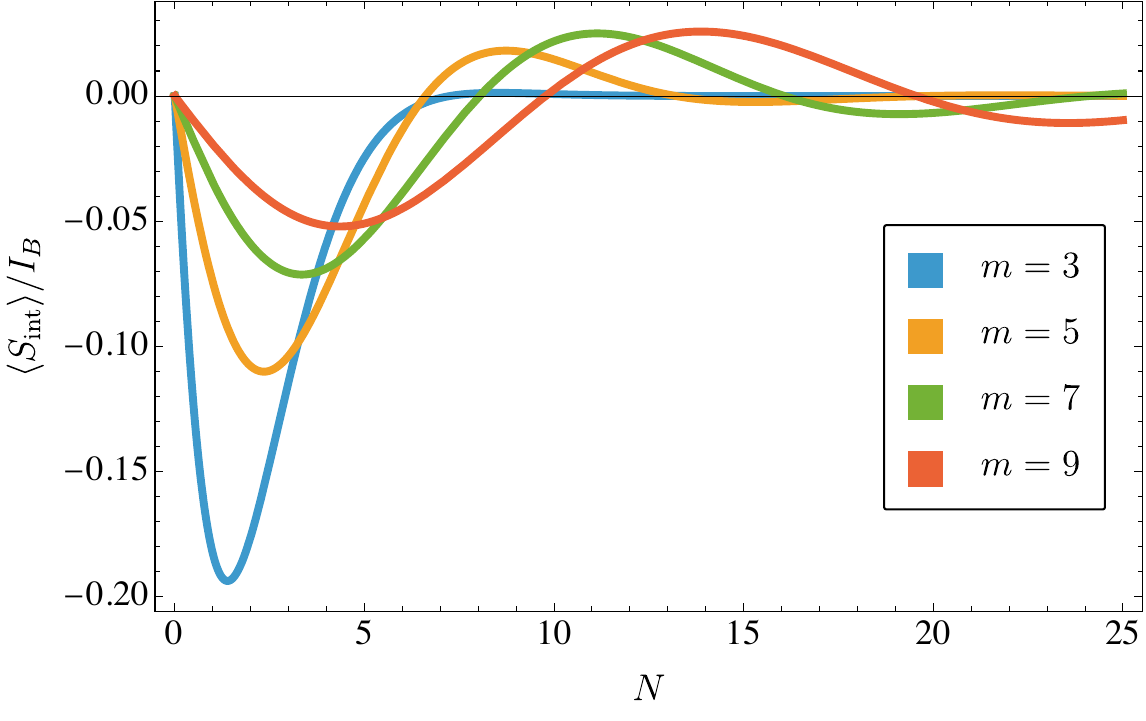}
   \caption{Aharonov--Bohm (AB) oscillations of the interference contribution to the tunneling current noise, $\langle S_{\rm int}\rangle$, shown as a function of $N = I_+ L_+ /(v e^\ast)$ for $e\lambda = e^\ast = e/m$. The parameters are $I_B = |\zeta_{\rm L}|\,|\zeta_{\rm R}|\,\tau_c^{2\lambda} I_+^{2\lambda-1}$, $\Phi = \pi/2$, and $I_- = 0$. The oscillation frequency is $\sin(2\pi\lambda)/2$.}
\label{fig:Sint_AB_oscillations}
\end{figure} We show the variation of $\langle S_{\rm int} \rangle$ against $N$ at $I_-=0$ in Fig.~\ref{fig:Sint_AB_oscillations}.

\subsection{Fano factor}

\add{We now turn to the second main result, which can help us measure the phase $\pi\lambda$ as a direct consequence of time-domain braiding at the interferometer QPCs.}

Let us examine the physical origin of the two terms appearing on the right-hand side of Eq.~\eqref{eq:S_int_Iminus0}. A comparison of Eqs.~\eqref{eq:interference_current_main} and~\eqref{eq:interference_noise} shows that the noise, which is defined through an anticommutator, contains an additional contribution arising from the $W$ integral [see Eq.~\eqref{eq: W(z)_definition}]. This contribution is absent in the tunneling current, where it cancels exactly. Such a cancellation is not accidental, but rather a characteristic feature of an FPI.

\begin{figure}
        \centering
    \includegraphics[width=0.80\linewidth]{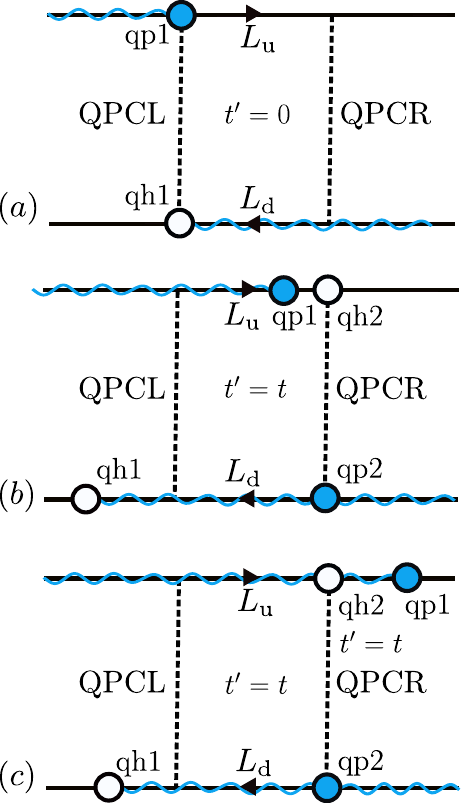}
    \caption{\add{Behavior of the commutator $[A^{\dagger}_{\rm L}(0), A_{\rm R}(t)]$
for $t$ both below and above $L_{\rm u}/v$ in an idealized geometry. Solid
lines denote the FPI arms, dashed lines denote QPCL and QPCR, and arrows
indicate the direction of QP and QH propagation. (a) At $t'=0$, $A^{\dagger}_{\rm L}(0)$ creates a QP--QH pair at
QPCL, with qp1 (blue) on edge u and qh1 (white) on edge d. Blue wiggly lines mark the co-moving region to their left. (b) The pair
propagates freely until $A_{\rm R}(t)$ creates a second pair at QPCR, with
qh2 on edge u and qp2 on edge d. For $t < L_{\rm u}/v$, qp1 has not yet
reached QPCR: the second pair forms inside the wiggly region of edge d but
outside that of edge u, the two arms contribute opposite phases
$e^{\pm i\pi\lambda}$, and the commutator vanishes. (c) Same as (b) but with
$t > L_{\rm u}/v$, so that qp1 has crossed QPCR by $t'=t$: qp1 and qh1 count both qh2 and qp2, the upper-arm phase flips
by $e^{-2\pi i\lambda}$, and
$[A^{\dagger}_{\rm L}(0), A_{\rm R}(t)] \neq 0$. This admits a topological interpretation: as qp1 crosses QPCR, its worldline becomes linked with that of qh1 at time $t$, and the resulting unit linking number is what renders the commutator nonvanishing.}}
    \label{fig:Fano_exchange_idealized}
\end{figure}

\add{Indeed, the nonlocal operator product $A^\dagger_{\rm L}(0)A_{\rm R}(t)$
and its reverse $A_{\rm R}(t)A^\dagger_{\rm L}(0)$ are related by
\begin{align} \label{locality_statement_2}
    A_{\rm L}^\dagger (0)A_{\rm R}(t) =&\, A_{\rm R}(t)A_{\rm L}^\dagger(0)\,
    \exp\bigl\{i\pi\lambda \bigl(\text{sign}[t+L_{\rm d}/v] \\ \nonumber
    &+\text{sign}[t-L_{\rm u}/v]\bigr)\bigr\}.
\end{align}
This expression implies that the commutator $[A^\dagger_{\rm L}(0),A_{\rm R}(t)]$
vanishes within the time window $t\in\left(-L_{\rm d}/v,\,L_{\rm u}/v\right)$,
as required by locality; see Sec.~\ref{Subsec:Bosonization} and the discussion
of Klein factors therein. This property is inherited directly
from the definition of the anyonic field operator in
Eq.~\eqref{definition of quasiparticle field operator} and the commutation
relation in Eq.~\eqref{bosoniccommutationrelation_in_time}. Moreover, its effect can be linked to time-domain braiding of QP--QH pairs created at QPCL and QPCR at different times. We illustrate this connection between locality and time-domain braiding in
Fig.~\ref{fig:Fano_exchange_idealized} for $t>0$; the case $t<0$ follows
analogously.}

\add{Figure~\ref{fig:Fano_exchange_idealized}(a) shows the FPI at time
$t'=0$, when the operator $A_{\rm L}^\dagger(0)$ creates a QP--QH pair
(qp1--qh1) at QPCL, with qp1 on edge u and qh1 on edge d. The pair propagates
freely for a time $t$, at which point $A_{\rm R}(t)$ creates a second pair
(qp2--qh2) at QPCR. Whether $A_{\rm L}^\dagger(0)A_{\rm R}(t)$ equals
$A_{\rm R}(t)A_{\rm L}^\dagger(0)$ then depends on whether $t$ was sufficient
for qp1 to cross QPCR along the upper arm. To understand this, we rewrite
Eq.~\eqref{locality_statement_2} as two separate identities for the two arms.
Denoting the u- and d-edge factors of the pair-creation operator
$A^{\dagger}_{\rm L}(0)$ by
$Q_{\rm u} \equiv e^{-i\phi_{\rm u}(x^{\rm L}_{\rm u},0)}$ and
$Q_{\rm d} \equiv e^{i\phi_{\rm d}(x^{\rm L}_{\rm d},0)}$,
we find
\begin{subequations} \label{eq:charge_counting_equations}
\begin{align}
\label{eq:charge_counting_equation_a}
Q^{\dagger}_{\rm u}\, e^{i\phi_{\rm u}(x^{\rm R}_{\rm u},t)}\, Q_{\rm u}
&= e^{i\phi_{\rm u}(x^{\rm R}_{\rm u},t)
}e^{i\pi\lambda\, \text{sign}\left(L_{\rm u}-vt\right)}, \\
\label{eq:charge_counting_equation_b}
Q^{\dagger}_{\rm d}\, e^{-i\phi_{\rm d}(x^{\rm R}_{\rm d},t)}\, Q_{\rm d}
&= e^{-i\phi_{\rm d}(x^{\rm R}_{\rm d},t)}
e^{-i \pi\lambda\, \text{sign}\left(L_{\rm d}+vt\right)}.
\end{align}
\end{subequations} The physical content of these
identities is that of fractional charge counting. Integrating
Eq.~\eqref{eq:charge_density} gives, up to a zero mode,
\begin{equation} \label{eq:counting_operator}
\phi_\alpha(x,t') = \frac{2\pi}{e}\int_{-\infty}^{x} dx'\,
\rho_\alpha(x',t'),
\end{equation}
so the vertex operators $e^{\pm i\phi_\alpha(x,t')}$ multiply a state by a
phase of $2\pi$ per electron charge located to the left of point $x$. Such an operator is blind to electrons, but registers each Laughlin QP (QH)
through the fractional phase shift $\pm 2\pi\lambda$: it acts as a fractional
charge \enquote{counter}, sensitive to the charge modulo $e$, i.e.,
to the QP or QH number modulo $1/\lambda$. Conversely, qp1 and qh1 can be viewed as fractional charge counters, counting all charge to their left (the region we represent with blue wiggly lines in 
Fig.~\ref{fig:Fano_exchange_idealized}) as they move along the FPI arms.}
\add{These lines give Eq.~\eqref{locality_statement_2} a direct pictorial meaning: the commutator $[A_{\rm L}^\dagger(0),A_{\rm R}(t)]$ is nonzero only if \textit{both} qp1 and qh1 can detect qp2 and qh2 at time $t$, i.e., when QPCR lies in the wiggly region along both the FPI arms when $A_{\rm R}(t)$ creates the second pair. In
Fig.~\ref{fig:Fano_exchange_idealized}(b), QPCR lies inside this region
for edge d, but outside it for edge u, since $t<L_{\rm u}/v$ and qp1 has not
yet reached QPCR. The two arms then contribute opposite exchange phases [see Eq.~\eqref{eq:charge_counting_equations}],
$e^{i\pi\lambda}$ and $e^{-i\pi\lambda}$, which cancel, and the commutator
vanishes. In Fig.~\ref{fig:Fano_exchange_idealized}(c), by contrast, qp1 has
swept across QPCR, so the second pair is created inside the wiggly regions of both arms. As a result, the phase contribution through the upper FPI arm flips, and we get a nonvanishing commutator. We stress that the phase contributions from the two FPI arms, which decide whether or not $[A_{\rm L}^\dagger(0),A_{\rm R}(t)]$ vanishes, can equally be understood through the lens of time-domain braiding. Indeed, the physics of Eq.~\eqref{eq:charge_counting_equations} is identical to that of Eq.~\eqref{bosonic_field_shift}: both describe a shift of the bosonic field $\phi_\alpha$ upon the creation of a QP (or QH) on edge $\alpha$. Whereas we previously used Eq.~\eqref{bosonic_field_shift} to describe a temporal interference process between an injected QP and a QP--QH pair created at an interferometer QPC, in the present context the role of the injected particle is played by qp1 on the upper arm and by qh1 on the lower one. In the language of geometric topology, the net phase contribution to the right-hand side of Eq.~\eqref{locality_statement_2} in both Fig.~\ref{fig:Fano_exchange_idealized}(b) and Fig.~\ref{fig:Fano_exchange_idealized}(c) is given by the linking number of the worldlines traversed by qp1 and qh1 between times $t'=0$ and $t'=t$~\cite{Ruelle2025,Polyakov1988,Witten1989,RevModPhys.80.1083,Han2016,Lee2022}, which changes from $0$ to $1$ as qp1 crosses QPCR at time $t$.}

\add{Since the tunneling current in Eq.~\eqref{eq:interference_current_main} is the
time integral of the above-mentioned commutator, the contribution of the window
$t\in(-L_{\rm d}/v,\,L_{\rm u}/v)$ cancels exactly [in
App.~\ref{App:Iint and Sint appendix}, this is the statement $I_{12}=I_{22}$], which
is why the $W$ integral of Eq.~\eqref{eq: W(z)_definition} is absent from
$\langle I_{\rm int}\rangle$. The noise in Eq.~\eqref{eq:interference_noise},
being defined through an anticommutator, instead adds the two operator
orderings as local fluctuations at QPCL and QPCR, and therefore receives
contributions from all times.} 
\add{The constructive addition of braiding phases corresponding to Fig.~\ref{fig:Fano_exchange_idealized}(c) enters $\langle I_{\rm int}\rangle$ and $\langle S_{\rm int} \rangle$ through the prefactors $\sin(\pi\lambda)$ in Eq.~\eqref{eq:interference_current_main} and $\cos(\pi\lambda)$ in the first term of Eq.~\eqref{eq:S_int_Iminus0}, whereas no such prefactor appears in its second term. In principle, it is possible to extract this additional phase by defining a ratio involving the second and first terms of Eq.~\eqref{eq:S_int_Iminus0}. To this end, we introduce an effective Fano factor~\cite{Rosenow,PhysRevLett.132.216601}, defined as the ratio
\begin{equation}
    P\!\left(\frac{I_-}{I_+},N\right)
    =
    \frac{
        \langle K_{\rm int}\rangle
    }{
        e^\ast I_+
        \left.
        \dfrac{\partial}{\partial I_-}
        \langle I_{\rm int} \rangle
        \right|_{I_-=0}
    } ,
    \label{eq:fano_factor_definition}
\end{equation}
where the denominator corresponds to the second term on the right-hand side of Eq.~\eqref{eq:cross_correlation_equation_int} evaluated at $I_-=0$. By construction, $P$ is dimensionless and therefore independent of microscopic parameters such as the backscattering amplitudes $\zeta_{\rm L,R}$ and the short-time cutoff $\tau_c$. Using the asymptotic
behavior $f(z)\propto z^{\lambda-1}e^{-z}$ for $z\to\infty$ ~\cite{DLMF} [see Eq.~\eqref{f_z_definition_main_text}],  we find that in the limit $I_-=0$ and large injection, i.e., $N\to\infty$, the Fano factor reduces to
\begin{equation}
\label{fano_factor_phase_shift}
    P(0,N \to \infty)
    =
    1
    -
    \frac{
        \sqrt{\pi}
        \bigl(N\sin^2(\pi\lambda)\bigr)^{-\lambda}
    }{
        2^{1-\lambda}
        \Gamma\!\left(\frac{3}{2}-\lambda\right)
    }
    \frac{
        \cos\!\left(\Phi_{\rm eff}\right)
    }{
        \cos\!\left(\Phi_{\rm eff}+\pi\lambda\right)
    }.
\end{equation}}

\begin{figure}
    \centering
   \includegraphics[width=1\linewidth]{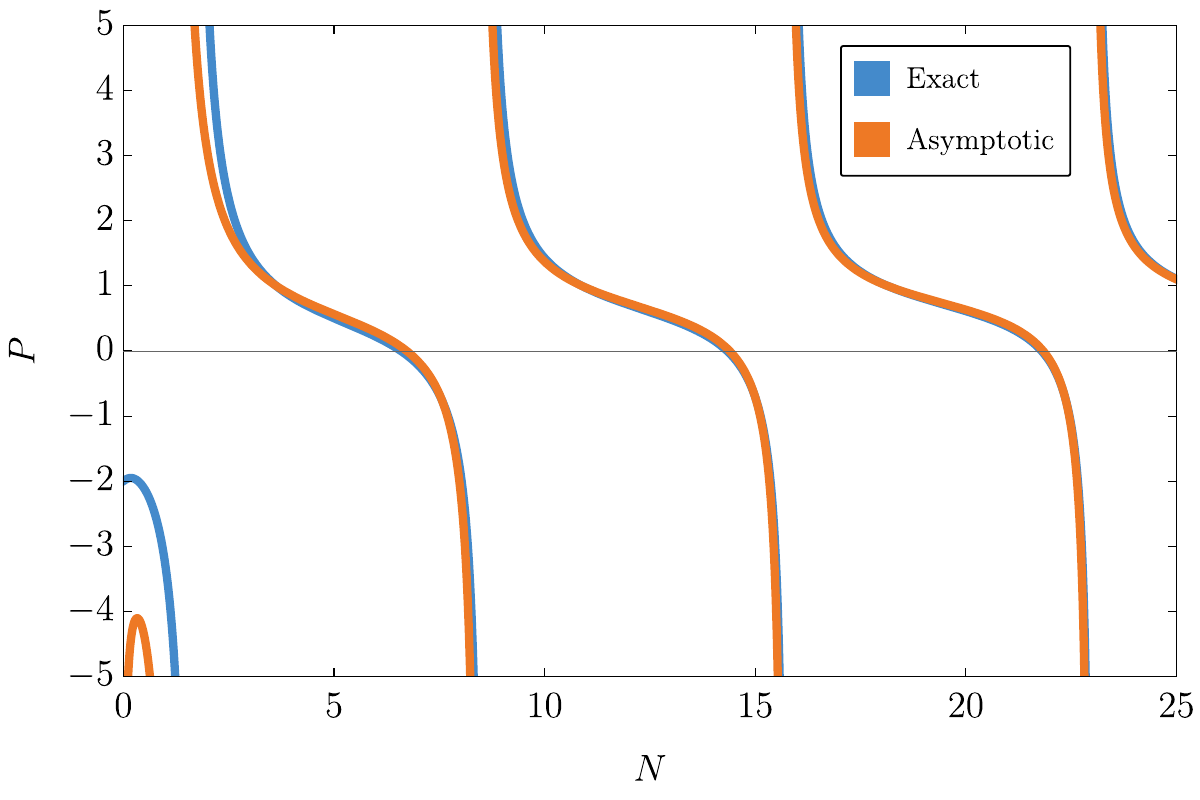}
\caption{\add{A comparison of the exact Fano factor in Eq.~\eqref{eq:fano_factor_definition} at $I_-=0$ to its asymptotic value in Eq.~\eqref{fano_factor_phase_shift} for the Laughlin state $e^\ast =e\lambda=e/3$. The bare AB phase $\Phi$ has been set to zero. We observe that already at $N \approx 5$, both expressions give almost identical results.}}
\label{fig:FanoComparison}
\end{figure}

 \add{As discussed, the phase shift of $\pi\lambda$ appearing in the denominator originates from time-domain braiding of QPs at the interferometer QPCs. Similar to the AB oscillations discussed in Sec.~\ref{zero_bias_noise}, Eq.~\eqref{fano_factor_phase_shift} thus also provides an experimentally accessible signature of the anyonic exchange phase $\pi\lambda$.}
 
\add{To illustrate the relevance of this result, we compare the exact and
asymptotic forms of the Fano factor
[Eqs.~\eqref{eq:fano_factor_definition} and
\eqref{fano_factor_phase_shift}] in Fig.~\ref{fig:FanoComparison}. Using
typical parameters from recent Fabry--P\'erot experiments (arm lengths
$\sim 1$--$3\ \mu$m, injected currents $\sim 50$--$200\ $pA chosen to
maintain the dilute Poissonian regime, and edge velocities
$v \sim 10^4$--$10^5\ $m/s)~\cite{nakamura2023fabryperot,Ofek2010,McClure2009,
Deprez2021}, we estimate occupations of $N \approx 0.5$--$1$ per arm, i.e.,
a total $N$ of order unity. Our numerical checks show that the asymptotic
large-$N$ expression [Eq.~\eqref{fano_factor_phase_shift}] becomes
essentially exact already for $N \gtrsim 5$. Since $N = I_+L_+/(ve^\ast)$ scales linearly with both the injected current and the interferometer
perimeter, this regime might be accessible in experimental devices and
could be reached by moderately increasing $I_+$ within the dilute limit, or
with slightly larger interferometers or slower edge modes. Signatures of the braiding phase $\pi\lambda$ should therefore be
observable in state-of-the-art or near-term devices, without requiring
unrealistically large interferometers or currents that would degrade
coherence.}

\section{Conclusion}
\label{Sec:Conclusion}
In this work, we have studied the interference of Laughlin quasiparticles (QPs) injected into a Fabry--Pérot interferometer (FPI) via Poissonian injection at two source contacts. We showed that due to time-domain braiding at the interferometer QPCs, the effective Aharonov--Bohm (AB) phase acquired by the QPs along the FPI loop includes an additional contribution of $\sin(2\pi\lambda)/2$ per QP, where $\pi\lambda$ is the QP exchange phase. For unequal incoming currents, an effective bias produces a net tunneling current between the interferometer arms. We computed the tunneling current, its noise, and the cross-correlator of the outgoing currents, and found that for equal incoming currents, the tunneling current noise exhibits purely AB oscillations as a function of the total number of QPs on the FPI loop, with frequency $\sin(2\pi\lambda)/2$. Moreover, we introduced a Fano factor, a universal, dimensionless quantity that captures the phase shift associated with \add{QP braiding at the interferometer QPCs.} Together, these results provide a simple route to measure the anyonic exchange phase $\pi\lambda$ via noise measurements.

Our model relies on a number of simplifications. We treat the edges as strictly one-dimensional and perfectly chiral, and the injections are assumed to be instantaneous and uncorrelated. Finite edge width, residual interactions, or extended injection pulses could affect the quantitative details of the results~\cite{PhysRevLett.132.216601,PhysRevB.83.155440,PhysRevLett.98.106801,feldman_robustness_2022}. Nevertheless, the overall structure of the tunneling current and its fluctuations is expected to remain qualitatively similar, and the framework we present can be adapted to explore more complex filling fractions.

In summary, our study provides a simple setting in which the interplay of injection, interference, and anyonic exchange can be analyzed using both tunneling current and noise as complementary probes of fractional statistics in quantum Hall systems.

\acknowledgments
We thank Kyrylo Snizhko,  Daniel Loss, Jelena Klinovaja and Sumathi Rao for fruitful discussions. S.G. gratefully acknowledges the financial support and warm hospitality of the University of Luxembourg, where this work was started. E.G.I. acknowledges financial support from the United Arab Emirates University under the Startup Grant No. G00004974, UPAR Grant No. G00005479, AUA-UAEU Grant No. G00005522 and SURE Plus Grant. T.L.S. acknowledges support by the QuantERA grant MAGMA from the National Research Fund Luxembourg under Grant No. INTER/QUANTERA21/16447820/MAGMA.

\appendix 
\begin{widetext}
\section{Derivation of \texorpdfstring{$\langle I_{\rm T} \rangle$}{<IT>} and \texorpdfstring{$\langle \delta I_T^2 \rangle_{\omega=0}$}{IT²{w=0}}}
\label{App:Iint and Sint appendix} \add{We start by calculating the generic contributions $\langle  I_{l l'} \rangle$ and $\langle  S_{l l'} \rangle,\ (l,l'\in \{\rm L,R\})$  to the tunneling current and its noise, respectively. We have
\begin{equation}\label{ILR_app}
\langle  I_{ll'} \rangle = e^{\ast}\int dt\ \left \langle \left[ A_{l}^\dagger(0),A_{l'}(t) \right] \right \rangle_0 = e^{\ast}\int dt\ \left(\langle A_{l}^\dagger(0) A_{l'}(t)\rangle_0 - \langle A_{l'}(t) A^\dagger_{l}(0)\rangle_0 \right)= e^{\ast} (I_1^{ll'} - I_2^{ll'})
\end{equation}  
and
\begin{equation}\label{SLR_app}
\langle  S_{ll'} \rangle = (e^{\ast})^2\int dt\ \left \langle \left\{ A_{l}^\dagger(0),A_{l'}(t) \right\} \right \rangle_0 = (e^{\ast})^2\int dt\ \left(\langle A_{l}^\dagger(0) A_{l'}(t)\rangle_0 + \langle A_{l'}(t) A^\dagger_{l}(0)\rangle_0 \right)= (e^{\ast})^2 (I_1^{ll'} + I_2^{ll'}),
\end{equation} where
  \begin{align}\label{I1_and_I2} 
    I_1^{ll'} 
&= 
    \int dt\ \zeta_{l}^{*} \zeta_{l'}\,
   \tau_c^{2\lambda}\,
\frac{e^{\frac{i \pi \lambda}{2} \left[\text{sign}
    \left(t_{\rm u}^{ll'}\right)+\text{sign}\left(t_{\rm d}^{ll'}\right) \right]}}{\left|t_{\rm u}^{ll'}  \right|^{\lambda}\left |t_{\rm d}^{ll'} \right|^{\lambda}}\chi_{\rm d} (t_{\rm d}^{ll'})
    \chi_{\rm u} (t_{\rm u}^{ll'}), \notag \\
    I_2^{ll'} 
&= \int dt\ \zeta_{l}^{*} \zeta_{l'}\,
   \tau_c^{2\lambda}\,
\frac{e^{-\frac{i \pi \lambda}{2} \left[\text{sign} \left(t_{\rm u}^{ll'}\right)+\text{sign}\left(t_{\rm d}^{ll'}\right) \right]}}{\left|t_{\rm u}^{ll'}  \right|^{\lambda}\left |t_{\rm d}^{ll'} \right|^{\lambda}}\chi_{\rm d} (t_{\rm d}^{ll'})
    \chi_{\rm u} (t_{\rm u}^{ll'}), \end{align} 
such that $t_{\rm u}^{ll'} \equiv t-(x^{l'}_{\rm u}-x^{l}_{\rm u})/v$, $t_{\rm d}^{ll'} \equiv t-(x^{l'}_{\rm d}-x^{l}_{\rm d})/v$ and $\chi_{\rm u/d}$ are given by Eqs.~\eqref{counting_statistics_generator_u} and~\eqref{counting_statistics_generator_d} in the main text.}
    
\subsection{Direct terms} \add{The direct terms $\langle I_{\rm dir}\rangle$ and $\langle S_{\rm dir} \rangle $ can be calculated when $l=l'={\rm L}$ and $l=l'=\rm R$. This translates to $t_{\rm u}^{\rm LL}=t_{\rm d}^{\rm LL}=t_{\rm u}^{\rm RR}=t_{\rm d}^{\rm RR}=t$. We find 
\begin{align}\label{ITdirect}
    \langle  I_{\rm dir} \rangle 
&= 
    C \left(|\zeta_{\rm L}|^2+ |\zeta_{\rm R}|^2\right)\sin(\pi\lambda) \int_{0}^{\infty} \frac{dt}{t^{2\lambda}} \frac{\sin \left ( \xi t \right )}{\exp \left (\Omega t\right)} \notag \\
&=
    C \bigl(|\zeta_{\rm L}|^2 + |\zeta_{\rm R}|^2\bigr)
    \Gamma(1-2\lambda)
    \left(\Omega^2+\xi^2\right)^{\lambda-\frac{1}{2}} \sin(\pi\lambda)\sin\!\left[(1-2\lambda)
    \tan^{-1}\!\left(\frac{\xi}{\Omega}\right)\right] \\
    \label{S_dir_app}
    \langle {S}_{\rm dir} \rangle 
&= 
    C' \left(|\zeta_{\rm L}|^2+ |\zeta_{\rm R}|^2\right)\cos(\pi\lambda) \int_{0}^{\infty} \frac{dt}{t^{2\lambda}} \frac{\cos \left ( \xi t \right )}{\exp \left (\Omega t\right)} \notag \\
&=
    C'
\left(|\zeta_{\rm L}|^2+|\zeta_{\rm R}|^2\right)
\Gamma(1-2\lambda)
\left(\Omega^2+\xi^2\right)^{\lambda-\frac{1}{2}} 
\cos(\pi\lambda)\cos\!\left[(1-2\lambda)\tan^{-1}\!\left(\frac{\xi}{\Omega}\right)\right],
\end{align} 
where $\Omega$ and $\xi$ are given by Eq.~\eqref{eq:omega_xi_def} in the main text.}

\subsection{Interference terms}
\add{Since $\langle I_{\rm LR} \rangle=\langle I_{\rm RL} \rangle^*$ and $\langle S_{\rm LR} \rangle=\langle S_{\rm RL} \rangle^*$, in the following we only calculate $\langle I_{\rm LR} \rangle$ and $\langle S_{\rm LR}\rangle$. This can be done by setting $l=\rm L$ and $l'=\rm R$ leading to $t_{\rm u}^{\rm LR}=t-L_{\rm u}/v$ and $t_{\rm d}^{\rm LR}=t+L_{\rm d}/v$ in Eq.~\eqref{I1_and_I2}. We decompose the integrals $I_1^{\rm LR}$ and $I_2^{\rm LR}$ into $I_n^{\rm LR}=I_{n1}+I_{n2}+I_{n3},\ n \in \{1,2\}$ such that  
\begin{equation}
    \label{I11} I_{11} = \int_{L_+ / v}^{\infty} dt \frac{\zeta_{\rm R}
    \zeta_{\rm L}^{\ast} \tau_c^{2 \lambda} e^{i \pi
    \lambda}}{(t - L_+ / v)^{\lambda}
    (t)^{\lambda}} \chi_{\rm d} (t)
    \chi_{\rm u} (t - L_+ / v),
\end{equation}
\begin{equation} \label{I12}
    I_{12} = \int_0^{L_+ / v} dt \frac{\zeta_{\rm R} \zeta_{\rm L}^{\ast} \tau_c^{2
    \lambda}}{(- t + L_+
    / v)^{\lambda} {(t)^{\lambda}}} \chi_{\rm d} ( t) \chi_{\rm u} (t - L_+ / v),
\end{equation}
\begin{equation} \label{I13}
    I_{13} = \int_0^{\infty} dt \frac{\zeta_{\rm R} \zeta_{\rm L}^{\ast} \tau_c^{2 \lambda}
   {e^{- i \pi \lambda}}}{(t + L_+ /
    v)^{\lambda} (t)^{\lambda}} \chi_{\rm d} (-t) \chi_{\rm u} (-t - L_+/v),
  \end{equation} and $I_{21}=e^{-2\pi i \lambda} I_{11},\ I_{22}= I_{12}$ and $I_{23}=e^{2 \pi i
    \lambda} I_{13}$. Each of the above three integrals can be calculated as follows.}

\subsubsection{\texorpdfstring{Evaluation of $I_{11}$}{Evaluation of I{11}}}
\begin{equation} 
    I_{11} = \zeta_{\rm R} \zeta_{\rm L}^{\ast} \tau_c^{2 \lambda} {e^{i
    \pi \lambda}} \int_{L_+ / v}^{\infty} dt \frac{\exp \left( {-
    \frac{\langle I_{{\rm d},0} \rangle t}{e^{\ast}} (1 - e^{2 \pi i \lambda})} -
    {\frac{\langle I_{{\rm u},0} \rangle (t - L_+ / v)}{e^{\ast}}
    (1 - e^{-2 \pi i \lambda})} \right)}{(t - L_+ /
    v)^{\lambda}(t)^{\lambda}}.
  \end{equation} Introducing the dimensionless integration variable $u = -1+\left(2vt/L_+\right)$, we get 
    \begin{equation} 
    I_{11} = |\zeta_{\rm R}| |\zeta_{\rm L}| \tau_c^{2 \lambda} e^{i
    \pi \lambda}\left( \frac{L_+}{2 v} \right)^{1 - 2
    \lambda}e^{i\Phi_{\rm eff}}e^{\mu}  f(z),
    \label{I11finalform}
  \end{equation} such that      \begin{equation} \label{f(z)_definition}f(z)=\int_1^{\infty} du\ \frac{e^{- u z}}{{(u - 1)^{\lambda}} {(u + 1)^{\lambda}}}=\frac{2^{\frac{1}{2}-\lambda} z^{\lambda-\frac{1}{2}} \Gamma (1-\lambda) K_{\frac{1}{2}-\lambda}(z)}{\sqrt{\pi }},  \end{equation} and $K_n(z)$ is the modified Bessel function of the second kind. The  parameters $\Phi_{\rm eff},\ \mu$ and $z$ can be read off from Eqs.~\eqref{mu_and_z_definition} and \eqref{eq:phi_effective_main_text_definition} in the main text.

\subsubsection{\texorpdfstring{Evaluation of $I_{12}$}{Evaluation of I{12}}}
  \begin{equation} 
    I_{12} =\zeta_{\rm R} \zeta_{\rm L}^{\ast} \tau_c^{2
    \lambda} \int_0^{L_+ / v} dt \frac{\exp \left( {-
    \frac{\langle I_{{\rm d},0} \rangle t}{e^{\ast}} (1 - e^{2 \pi i \lambda})} +
    {\frac{\langle I_{{\rm u},0} \rangle (t - L_+ / v)}{e^{\ast}}
    (1 - e^{2 \pi i \lambda})} \right)}{{(- t + L_+
    / v)^{\lambda}} {(t)^{\lambda}}}.
  \end{equation} Again substituting $u=-1+(2vt/L_+)$ gives
  \begin{equation}
    \label{I12finalform} I_{12} = |\zeta_{\rm R}| |\zeta_{\rm L}| \tau_c^{2 \lambda}
   \left( \frac{L_+}{2 v} \right)^{1 - 2
    \lambda}e^{i\Phi_{\rm eff}} e^{-\frac{L_+\Omega}{2v}} W (\tilde z),
  \end{equation} where \begin{equation} W(\tilde z)=\int_{-
    1}^1 du \frac{e^{u \tilde z}}{{(1 + u)^{\lambda}}
    {(1 - u)^{\lambda}}}=\sum_{k=0}^{\infty} \left(\frac{\tilde z}{2}\right)^{2k}\left(\frac{\sqrt{\pi}\ \Gamma(1-\lambda)}{k! \cdot \Gamma\left(k+\frac{3}{2}-\lambda\right)}\right),\end{equation} and $\tilde z$ can be read off from Eq.~\eqref{tilde_z_definition}.
\subsubsection{\texorpdfstring{Evaluation of $I_{13}$}{Evaluation of I{13}}}
 The third integral is very similar to the first one. We have
\begin{equation} 
    I_{13} = \zeta_{\rm R} \zeta_{\rm L}^{\ast} \tau_c^{2 \lambda} {e^{-i
    \pi \lambda}} \int_{0}^{\infty} dt \frac{\exp \left( {-
    \frac{\langle I_{{\rm d},0} \rangle t}{e^{\ast}} (1 - e^{-2 \pi i \lambda})} -
    {\frac{\langle I_{{\rm u},0} \rangle (t + L_+ / v)}{e^{\ast}}
    (1 - e^{2 \pi i \lambda})} \right)}{(t + L_+ /
    v)^{\lambda}(t)^{\lambda}}.
  \end{equation} After the substitution $u=1+(2vt/L_+)$, we get
  \begin{equation}
    \label{I13finalform} I_{13} = |\zeta_{\rm R}| |\zeta_{\rm L}| \tau_c^{2 \lambda} e^{-i\pi\lambda} \left( \frac{L_+}{2 v} \right)^{1 - 2 \lambda}e^{i\Phi_{\rm eff}}e^{-\mu}   f
    (z^*).
  \end{equation} 

\subsubsection{\texorpdfstring{Evaluation of $\langle I_{\rm int} \rangle$ and $\langle S_{\rm int} \rangle$}{Evaluation of Iint and Sint}} Using Eqs. \eqref{I11finalform}, \eqref{I12finalform} and \eqref{I13finalform}, we find 
\begin{equation}
\langle I_{\rm int}\rangle=2e^\ast\text{Re}\left(I_{11}-I_{21}+I_{13}-I_{23}\right)= 2e^\ast \text{Re}\left((1-e^{-2\pi i \lambda})I_{11}+(1-e^{2\pi i \lambda})I_{13}\right),
\end{equation} 
\begin{equation}
\langle S_{\rm int}\rangle=2(e^\ast)^2\ \text{Re}\left(I_{11}+I_{21}+2I_{22}+I_{13}+I_{23}\right)= 2(e^\ast)^2\text{Re}\left((1+e^{-2\pi i \lambda})I_{11}+(1+e^{2\pi i \lambda})I_{13}+2I_{12}\right).
\end{equation} These equations reduce to the corresponding results in the main text.

\section{Interference terms in cross-correlations \texorpdfstring{$\langle \delta I_{\rm u} \delta I_{\rm d} \rangle_{\omega=0}$}{cross-correlation noise at zero frequency}}
\label{Appendix:interference terms in cross-correlations}
 Using Eq.~\eqref{eq:cross_correlation_equation_int} from the main text, the interference part of $\langle \delta I_{\rm u} \delta I_{\rm d} \rangle_{\omega=0}$ is

\begin{equation}
    \label{K_int}
   \langle K_{\rm int} \rangle =2\text{Re}(\langle K_{LR} \rangle) =-\langle S_{\rm int} \rangle+ e^{\ast} \left (\langle  {I}_{\rm u,0} \rangle \frac{\partial }{\partial \langle {I}_{\rm u,0} \rangle}- \langle {I}_{\rm d,0} \rangle \frac{\partial }{\partial \langle {I}_{\rm d,0} \rangle} \right) \langle I_{\rm int} \rangle.
\end{equation} We can calculate the derivatives above by differentiating the integrals $I_{1}$ and $I_{2}$ in App.~\ref{App:Iint and Sint appendix} under the integral sign. For the sake of brevity, we introduce the notation $\left (\langle  {I}_{\rm u,0} \rangle \frac{\partial }{\partial \langle {I}_{\rm u,0} \rangle}- \langle {I}_{\rm d,0} \rangle \frac{\partial }{\partial \langle {I}_{\rm d,0} \rangle} \right) \equiv \hat \partial$. We find

\begin{equation} \label{I11derivative}
     e^\ast \hat\partial I_{11}=|\zeta_{\rm R}| |\zeta_{\rm L}| \tau_c^{2 \lambda} {e^{i
     \pi \lambda}}  \left(\frac{L_+}{2v}\right)^{2-2\lambda} e^{i\Phi_{\rm eff}}e^{\mu } \left\{\left[\langle I_{{\rm d},0} \rangle (1 - e^{2 \pi i \lambda}) \right] M_+(\lambda, z)-\left[\langle I_{{\rm u},0} \rangle (1 - e^{-2 \pi i \lambda}) \right] M_-(\lambda,z) \right\},
\end{equation} where the integral $M_{\pm}(\lambda,z)$ is defined as

\begin{equation}
        \label{Mintegral}
       M_{\pm}(\lambda,z)\equiv\int_{1}^{\infty} du \frac{(u \pm 1) \exp{(-z \cdot u)}}{(u+1)^\lambda (u-1)^\lambda} =\frac{2^{1/2-\lambda}\Gamma(1-\lambda)}{\sqrt{\pi}}z^{\lambda-1/2} \left (K_{\lambda-3/2}(z) \pm K_{\lambda-1/2}(z) \right).   \end{equation}  Similarly, we find

\begin{equation} \label{I13derivative}
    e^\ast \hat \partial I_{13}=|\zeta_{\rm R}| |\zeta_{\rm L}| \tau_c^{2 \lambda} {e^{-i
     \pi \lambda}}  \left(\frac{L_+}{2v}\right)^{2-2\lambda} e^{i\Phi_{\rm eff}}e^{-\mu }\left\{\left[\langle I_{{\rm d},0} \rangle (1 - e^{-2 \pi i \lambda}) \right] M_-(\lambda,z^*)-\left[\langle I_{{\rm u},0} \rangle (1 - e^{2 \pi i \lambda}) \right] M_+(\lambda,z^*) \right\}.
\end{equation} Using Eqs.~\eqref{I11derivative} and \eqref{I13derivative},
\begin{align} \label{ILRderivative}
   &e^\ast \hat \partial \langle I_{\rm LR} \rangle= 2ie^\ast \sin(\pi\lambda)|\zeta_{\rm R}| |\zeta_{\rm L}| \tau_c^{2 \lambda}  \left(\frac{L_+}{2v}\right)^{2-2\lambda}e^{i\Phi_{\rm eff}} \Bigg[e^{\mu} \bigg\{\left[\langle I_{{\rm d},0} \rangle (1 - e^{2 \pi i \lambda}) \right] M_+(\lambda,z)-\left[\langle I_{{\rm u},0} \rangle (1 - e^{-2 \pi i \lambda}) \right] \notag \\
     &\quad  M_-(\lambda,z) \bigg\} -e^{-\mu }\bigg\{\left[\langle I_{{\rm d},0} \rangle (1 - e^{-2 \pi i \lambda}) \right] M_-(\lambda,z^*)-\left[\langle I_{{\rm u},0} \rangle (1 - e^{2 \pi i \lambda}) \right] M_+(\lambda,z^*) \bigg\}\Bigg]. 
\end{align}
The above result can be used to calculate $\langle K_{\rm int} \rangle =-\langle S_{\rm int}\rangle +2e^\ast\text{Re}\left(\hat \partial \langle  I_{\rm LR} \rangle\right)$.
    
\section{Derivation of the cross-correlator \texorpdfstring{$\langle \delta I_{\rm u}\delta I_{\rm d} \rangle_{\omega=0}$}{δ Iu δ Id{ω=0}}}
\label{Appendix:cross_correlation_derivation}
The cross term $\langle \delta { I_{\rm T}} \delta  I_{{\rm u},0}\rangle_{\omega=0}$ is given by
\begin{equation}
\langle \delta { I_{\rm T}} \delta  I_{{\rm u},0}\rangle_{\omega=0}=\int_{-\infty}^{+\infty} dt\ \langle { I_{\rm T}}(0) \delta  I_{{\rm u},0}(t)\rangle=-e^{\ast}\sum_{l,l^{\prime}} \int_{-\infty}^{+\infty} dt \int_{-\infty}^0 dt^{\prime}\langle [A_l(t^{\prime})+A^{\dagger}_l(t^{\prime}), A_{l^{\prime}}(0)-A^{\dagger}_{l^{\prime}}(0)]\delta I_{{\rm u},0} (t)\rangle.
\end{equation}
Expanding the commutator, we get 
\begin{equation}\label{a2}	
\langle \delta { I_{\rm T}} \delta  I_{{\rm u},0}\rangle_{\omega=0}=   \left(P_1+P_2+P_3+P_4\right),
\end{equation}
with the four terms given by
\begin{equation}
\label{a3}	
\begin{split}
& P_1=-e^{\ast}\sum_{l,l^\prime }\int^0_{-\infty}dt^{\prime}  \int dt \langle A^{\dagger}_l(t^{\prime}) A_{l^{\prime}}(0)\delta  I_{{\rm u},0}(t)\rangle, \quad P_2=e^{\ast} \sum_{l,l^\prime }\int^0_{-\infty}dt^{\prime} \int dt \langle A_{l^{\prime}}(0) A_l^{\dagger}(t^{\prime})\delta  I_{{\rm u},0}(t)\rangle, \\
& P_3=-e^{\ast} \sum_{l,l^\prime } \int^0_{-\infty}dt^{\prime} \int dt \langle A^{\dagger}_{l^{\prime}}(0) A_{l}(t^{\prime})\delta  I_{{\rm u},0}(t)\rangle, \quad P_4=e^{\ast} \sum_{l, l^\prime }\int^0_{-\infty}dt^{\prime} \int dt \langle A_l(t^{\prime}) A^{\dagger}_{l^{\prime}}(0)\delta  I_{{\rm u},0}(t)\rangle.
\end{split}
\end{equation} 
On the other hand, the tunneling current is also given by four terms, see Eq.~\eqref{currentexpectationvalue} in the main text:
\begin{equation}
\label{a4}	
\langle I_{\rm T} \rangle= -e^{\ast} \sum_{l,l^\prime } \int_{-\infty}^{0}dt' \left(\langle A_l^\dagger (t') A_{l'}(0) \rangle - \langle A_{l'}(0) A_l^\dagger (t') \rangle + \langle A_{l'}^\dagger (0) A_{l}(t') \rangle - \langle A_l(t')A_{l'}^\dagger(0) \rangle\right), 
\end{equation}
which we call $I_{p_1},\ I_{p_2},\ I_{p_3}$ and $I_{p_4}$ respectively. We now want to write $P_1$ in terms of the derivative of $I_{p_1}$ with respect to $\langle  I_{{\rm u},0} \rangle$. We start by introducing the greater Green function 
 \begin{equation}
 \label{Green_functions}
iG^>_{d} (t)= \langle e^{i \phi_{\rm d}(0,t)} e^{-i\phi_{\rm d} (0,0)} \rangle.
\end{equation} Writing $I_{p_1}$ in terms of the bosonic fields $\phi_{\rm u}(x,t)$ and $\phi_{\rm d}(x,t)$ gives
\begin{equation}
\label{a8}
I_{p_1} = 
-e^{*} \sum_{l,l'} \zeta_{l}^{*} \zeta_{l'} \int_{-\infty}^{0} dt' \,
\left[ i G^{>}_{d} \left( t' - \frac{x^{l}_{\rm d} - x^{l'}_{\rm d}}{v} \right) \right]\left\langle e^{-i \phi_{\rm u} \left( 0, t' - x_{\rm u}^l / v \right)} \, 
e^{i \phi_{\rm u} \left( 0, -x_{\rm u}^{l'} / v \right)} \right\rangle_{\text{eq.}}\ \chi_{\rm u} \left( -\left( t' - \frac{x^l_{\rm u} - x^{l'}_{\rm u}}{v} \right) \right).
\end{equation}
Under the substitution $t'-\left(x^{l}_{ \rm u}-x^{l'}_{\rm u}\right)/v=\tilde t$, we find 
\begin{equation}
\label{a9}
I_{p_1} = -e^{*} \sum_{l,l'} \zeta_{l}^{*} \zeta_{l'} \int_{-\infty}^{\left( x_{\rm u}^{l'} - x_{\rm u}^l \right)/v} d\tilde{t} \,
\left[ i G^{>}_{d} \left( \tilde{t} + \frac{x^{l}_{\rm u} - x^{l'}_{\rm u}}{v} - \frac{x^{l}_{\rm d} - x^{l'}_{\rm d}}{v} \right) \right]
\left\langle e^{-i \phi_{\rm u}(0,\tilde{t})} \, e^{i \phi_{\rm u}(0,0)} \right\rangle_{\text{eq.}} \, \chi_{\rm u}(-\tilde{t}).
\end{equation}
Now consider $P_1$,
\begin{equation}
\label{a10}
P_1=-e^{\ast}\sum_{l,l'}\int_{-\infty}^{0}\ dt' \left \langle A_l^\dagger (t') A_{l'}(0) \int dt\ \delta  I_{{\rm u},0} (t) \right \rangle. 
\end{equation} We define $e^\ast \delta  N_{{\rm u},0}(t) \equiv \delta I_{{\rm u},0}(t)dt$ as the average fluctuations in the injected QP charge via QPCU from time $t$ to $t+dt$. Then the nonequilibrium bosonic field can be decomposed as $\phi_{\rm u}(x,t)=\phi_{\rm u}^{\rm eq.}(x,t)+2\pi\lambda N_{{\rm u},0}(t)$, where $N_{{\rm u},0}(t)$ is a \enquote{counting variable} with $\langle N_{{\rm u},0}(t)\rangle-\langle N_{{\rm u},0}(0) \rangle=\langle I_{{\rm u},0} \rangle t/e^\ast$ [see Eq.~\eqref{bosonic_field_shift} in the main text] and is uncorrelated with the equilibrium field $\phi_{\rm u}^{\rm eq.}$. Using $N_{{\rm u},0}(t)=\langle N_{{\rm u},0}(t) \rangle + \delta N_{{\rm u},0} (t)$, we find
\begin{equation} \label{a11}
\begin{split}	
P_1 = -e^{\ast} \sum_{l,l'} \int_{-\infty}^{\left( x_{\rm u}^{l'} - x_{\rm u}^l \right)/v} d\tilde{t} \ 
\zeta^{\ast}_l \zeta_{l'} \left[ i G^>_{d} \left( \tilde{t} + \frac{x^l_{\rm u} - x^{l'}_{\rm u}}{v} - \frac{x^l_{\rm d} - x^{l'}_{\rm d}}{v} \right) \right] 
\left\langle e^{-i \phi_{\rm u}(0,\tilde{t})} e^{i \phi_{\rm u}(0,0)} \right\rangle_{\text{eq.}}
\\
 \cdot e^{\ast} \left\langle e^{-2\pi i \lambda \langle I_{{\rm u},0} \rangle \tilde{t} / e^\ast } \
e^{2\pi i \lambda \left[ \delta N_{{\rm u},0}(-\tilde{t}) - \delta N_{{\rm u},0}(0) \right]} 
\int_{-\infty}^{+\infty} dt \, \partial_t \delta N_{{\rm u},0}(t) \right\rangle.
\end{split}	
\end{equation} where we used $\left \langle e^{(2\pi i \lambda) \left(N_{{\rm u},0}(-\tilde t)-N_{{\rm u},0}(0)\right)} \right \rangle=\chi_{\rm u}(-\tilde t)$. Since the injection events are Markovian, the main contribution to the classical correlation function above will be nonzero only in the interval $t \in (\tilde t,\ 0)$. Its limits can then be truncated as $\int_{-\infty}^{+\infty} \equiv \text{sign} \left(-\tilde t\right) \int_{0}^{-\tilde t}$, where the $\text{sign}$ function preserves the sign of the time step $dt$ when the sign of $\tilde t$ changes. This gives us
\begin{equation} \label{a13}
\begin{split}
P_1 = 
&-e^{*} \sum_{l,l'} \int_{-\infty}^{\left( x_{\rm u}^{l'} - x_{\rm u}^l \right)/v} \, d\tilde{t} \, 
\zeta^{*}_l \zeta_{l'} \left[ i G^{>}_{d} \left( \tilde{t} + \frac{x^{l}_{\rm u} - x^{l'}_{\rm u}}{v} - \frac{x^{l}_{\rm d} - x^{l'}_{d}}{v} \right) \right] 
\left\langle e^{-i \phi_{\rm u}(0,\tilde{t})} \, e^{i \phi_{\rm u}(0,0)} \right\rangle_{\text{eq.}} \\
&\times e^{*} \, \operatorname{sign}(-\tilde{t}) \left\langle 
e^{-2\pi i \lambda \langle I_{{\rm u},0} \rangle \tilde{t} / e^{*}} 
\frac{\partial}{\partial (2\pi i \lambda)} 
e^{2\pi i \lambda \left[ \delta N_{{\rm u},0}(-\tilde{t}) - \delta N_{{\rm u},0}(0) \right]} 
\right\rangle \\
={}& 
-e^{*} \sum_{l,l'} \int_{-\infty}^{\left( x_{\rm u}^{l'} - x_{\rm u}^l \right)/v} \, d\tilde{t} \, 
\zeta^{*}_l \zeta_{l'} \left[ i G^{>}_{d} \left( \tilde{t} + \frac{x^{l}_{\rm u} - x^{l'}_{\rm u}}{v} - \frac{x^{l}_{\rm d} - x^{l'}_{\rm d}}{v} \right) \right] 
\left\langle e^{-i \phi_{\rm u}(0,\tilde{t})} \, e^{i \phi_{\rm u}(0,0)} \right\rangle_{\text{eq.}} \\
&\times e^{*} \, \operatorname{sign}(-\tilde{t}) \left\langle 
e^{-2\pi i \lambda \langle I_{{\rm u},0} \rangle \tilde{t} / e^{*}} 
\frac{\partial}{\partial (2\pi i \lambda)} 
\left[ \chi_{\rm u}(-\tilde{t}) \, e^{2\pi i \lambda \langle I_{{\rm u},0} \rangle \tilde{t} / e^{*}} \right] 
\right\rangle.
\end{split}
\end{equation} Recalling that 
\begin{equation}\label{a15}
 \chi_{\rm u}(t)=\exp \left[- |t| \frac{\langle  I_{{\rm u},0} \rangle}{e^{\ast}}\left(1-e^{-2\pi i\lambda\ \text{sign}(t)}\right)\right],
 \end{equation} the above can be rewritten as

\begin{equation}
\label{a16}
\begin{split}
P_1 = 
&-e^{*} \langle I_{{\rm u},0} \rangle \frac{\partial}{\partial \langle I_{{\rm u},0} \rangle} 
\sum_{l,l'} e^{*} \zeta^{*}_l \zeta_{l'} \int_{-\infty}^{\left( x_{\rm u}^{l'} - x_{\rm u}^l \right)/v} 
d\tilde{t} \, 
\left[ i G^{>}_{d} \left( \tilde{t} + \frac{x^{l}_{\rm u} - x^{l'}_{\rm u}}{v} - \frac{x^{l}_{\rm d} - x^{l'}_{\rm d}}{v} \right) \right] 
\left\langle e^{-i \phi_{\rm u}(0,\tilde{t})} \, e^{i \phi_{\rm u}(0,0)} \right\rangle_{\text{eq.}} \\
&\times \chi_{\rm u}(-\tilde{t}).
\end{split}
\end{equation}
Comparing this with Eq.~\eqref{a9} gives 
 \begin{equation}
 \label{a17}
 P_1=e^{\ast} \langle  I_{{\rm u},0} \rangle \frac{\partial }{\partial \langle  I_{{\rm u},0} \rangle} I_{p_1}. 
 \end{equation}
 In this same manner, we can show that
\begin{equation}
\label{a18}
P_2=e^{\ast} \langle  I_{{\rm u},0} \rangle \frac{\partial }{\partial \langle  I_{{\rm u},0} \rangle} I_{p_2}, \quad 
P_3=e^{\ast} \langle  I_{{\rm u},0} \rangle \frac{\partial }{\partial \langle  I_{{\rm u},0} \rangle} I_{p_3}, \quad P_4=e^{\ast} \langle  I_{{\rm u},0} \rangle \frac{\partial }{\partial \langle  I_{{\rm u},0} \rangle} I_{p_4}. 
\end{equation}
Adding Eqs.~\eqref{a17} and \eqref{a18} gives 
 \begin{equation}
 \label{a19}
 \langle  \delta { I_{\rm T}} \delta  I_{{\rm u},0}\rangle_{\omega=0} =e^\ast \langle  I_{{\rm u},0} \rangle \frac{\partial }{\partial \langle  I_{{\rm u},0} \rangle}  \langle { I_{\rm T}} \rangle. 
 \end{equation}
 Similarly, for the down term, we can show that 
 \begin{equation}\label{a20}
 \langle  \delta { I_{\rm T}} \delta  I_{{\rm d},0}\rangle_{\omega=0} =e^\ast \langle  I_{{\rm d},0} \rangle \frac{\partial }{\partial \langle  I_{{\rm d},0} \rangle}  \langle { I_{\rm T}} \rangle. 
 \end{equation}

\section{Generalization to \texorpdfstring{$n$}{n} QPCs} \label{n_QPCs}
We assume that the QPCs on the upper (lower) edge are located at positions $x^{j}_{\rm u,d}$, respectively, with 
\begin{align}
    & x^1_{\rm u} < x^2_{\rm u} < \ldots < x^n_{\rm u}, \notag \\
    & x^1_{\rm d} > x^2_{\rm d} > \ldots > x^n_{\rm d}.
\end{align}
Generalizing the setup in Fig.~\ref{fig:FPI_HallBar_Schematic}, we allow for QP tunneling with amplitudes $\zeta_j$ between positions $x^j_{\rm u}$ and $x^j_{\rm d}$. As discussed in Sec.~\ref{Sec: Model_and_Summary} in the main text, the limit of vanishing interferometer length, $L_+ \to 0$, corresponds to fully coherent interference. In this limit, the \emph{direct} tunneling current at a given QPC can be viewed as the \emph{interference} contribution arising from two QPCs separated by distance zero.

With this interpretation, the tunneling current for $n$ QPCs can be written in the form

\add{\begin{equation}\label{eq:interference_current_n_qpcs}
\langle I_T\rangle=-C\sum_{j\ge k}\Big(1-\tfrac{\delta_{jk}}{2}\Big)\sum_{\gamma=\pm1}\Big(\frac{L^{jk}_+}{2v}\Big)^{1-2\lambda}|\zeta_j||\zeta_k|\sin(\pi\lambda)\,
\mathrm{Im}\Big\{e^{i\Phi^{jk}_{\rm eff}}\,\gamma\,e^{\gamma\mu^{jk}}f\Big(\frac{L^{jk}_+}{2v}(\Omega+i\gamma\xi)\Big)\Big\},
\end{equation} where $\delta_{jk}$ is the Kronecker delta function and} $L^{jk}_+ = (x^j_{\rm u} - x^k_{\rm u}) - (x^j_{\rm d} - x^k_{\rm d})$ denotes the edge length associated with a loop connecting QPCs $j$ and $k$. The variables $\Phi_{\rm eff}^{jk}$ and $\mu^{jk}$ are defined analogously,
\begin{align}
    \Phi_{\rm eff}^{jk} =\Phi^{jk}+\frac{I_+L_+^{jk}}{2ve^\ast}\sin(2\pi\lambda),\  \mu^{jk} &= \frac{I_-L_+^{jk}}{2v e^\ast}\bigl[1-\cos(2\pi\lambda)\bigr].
\end{align} The corresponding results for $\langle \delta I_{\rm T}^2 \rangle_{\omega=0}$ and $\langle \delta I_{\rm u} \delta I_{\rm d}\rangle_{\omega=0}$ follow in a similar manner.
\end{widetext}

\bibliography{refs}

@article{Tsui,
  title = {Two-Dimensional Magnetotransport in the Extreme Quantum Limit},
  author = {Tsui, D. C. and Stormer, H. L. and Gossard, A. C.},
  journal = {Phys. Rev. Lett.},
  volume = {48},
  issue = {22},
  pages = {1559--1562},
  numpages = {0},
  year = {1982},
  month = {May},
  publisher = {American Physical Society},
  doi = {10.1103/PhysRevLett.48.1559},
  url = {https://link.aps.org/doi/10.1103/PhysRevLett.48.1559}
}

@article{HeiblumStern,
doi = {10.1088/2058-7058/13/3/30},
url = {https://dx.doi.org/10.1088/2058-7058/13/3/30},
year = {2000},
month = {mar},
publisher = {},
volume = {13},
number = {3},
pages = {37},
author = {Moty Heiblum and Ady Stern},
title = {Fractional quantum {Hall} effects},
journal = {Physics World}}

@article{Laughlin,
  title = {Anomalous Quantum {Hall} Effect: An Incompressible Quantum Fluid with Fractionally Charged Excitations},
  author = {Laughlin, R. B.},
  journal = {Phys. Rev. Lett.},
  volume = {50},
  issue = {18},
  pages = {1395--1398},
  numpages = {0},
  year = {1983},
  month = {May},
  publisher = {American Physical Society},
  doi = {10.1103/PhysRevLett.50.1395},
  url = {https://link.aps.org/doi/10.1103/PhysRevLett.50.1395}
}

@book{Halperin2020,
  editor  = {Halperin, Bertrand I. and Jain, Jainendra K.},
  title   = {Fractional Quantum {Hall} Effects: New Developments},
  publisher = {World Scientific},
  year    = {2020},
  doi     = {10.1142/11751},
  isbn    = {978-981-121-748-7},
  url     = {https://www.worldscientific.com/worldscibooks/10.1142/11751}
}

@article{HeiblumFeldman,
author = {Heiblum, Moty and Feldman, D. E.},
title = {Edge probes of topological order},
journal = {Int. J. Mod. Phys. A},
volume = {35},
number = {18},
pages = {2030009},
year = {2020},
doi = {10.1142/S0217751X20300094},
URL = {https://doi.org/10.1142/S0217751X20300094}
}

@article{IdrisovSchmidt,
  title = {Current cross correlations in a quantum {Hall} collider at filling factor two},
  author = {Idrisov, Edvin G. and Levkivskyi, Ivan P. and Sukhorukov, Eugene V. and Schmidt, Thomas L.},
  journal = {Phys. Rev. B},
  volume = {106},
  issue = {8},
  pages = {085405},
  numpages = {6},
  year = {2022},
  month = {Aug},
  publisher = {American Physical Society},
  doi = {10.1103/PhysRevB.106.085405},
  url = {https://link.aps.org/doi/10.1103/PhysRevB.106.085405}
}

@article{Kane1,
  title = {Randomness at the edge: Theory of quantum {Hall} transport at filling \ensuremath{\nu}=2/3},
  author = {Kane, C. L. and Fisher, Matthew P. A. and Polchinski, J.},
  journal = {Phys. Rev. Lett.},
  volume = {72},
  issue = {26},
  pages = {4129--4132},
  numpages = {0},
  year = {1994},
  month = {Jun},
  publisher = {American Physical Society},
  doi = {10.1103/PhysRevLett.72.4129},
  url = {https://link.aps.org/doi/10.1103/PhysRevLett.72.4129}
}

@article{Kane2,
  title = {Impurity scattering and transport of fractional quantum {Hall} edge states},
  author = {Kane, C. L. and Fisher, Matthew P. A.},
  journal = {Phys. Rev. B},
  volume = {51},
  issue = {19},
  pages = {13449--13466},
  numpages = {0},
  year = {1995},
  month = {May},
  publisher = {American Physical Society},
  doi = {10.1103/PhysRevB.51.13449},
  url = {https://link.aps.org/doi/10.1103/PhysRevB.51.13449}
}

@article{Mielke,
  author    = {Mielke, A.},
  title     = {Disorder and the fractional quantum {Hall} effect: the reduction of the gap},
  journal   = {Journal of Physics: Condensed Matter},
  volume    = {2},
  number    = {48},
  pages     = {9567},
  year      = {1990},
  doi       = {10.1088/0953-8984/2/48/010},
  publisher = {IOP Publishing},
}

@article{Mross,
  title = {Theory of Disorder-Induced Half-Integer Thermal {Hall} Conductance},
  author = {Mross, David F. and Oreg, Yuval and Stern, Ady and Margalit, Gilad and Heiblum, Moty},
  journal = {Phys. Rev. Lett.},
  volume = {121},
  issue = {2},
  pages = {026801},
  numpages = {6},
  year = {2018},
  month = {Jul},
  publisher = {American Physical Society},
  doi = {10.1103/PhysRevLett.121.026801},
  url = {https://link.aps.org/doi/10.1103/PhysRevLett.121.026801}
}

@article{Shytov1,
  title = {Tunneling into the Edge of a Compressible Quantum {Hall} State},
  author = {Shytov, A. V. and Levitov, L. S. and Halperin, B. I.},
  journal = {Phys. Rev. Lett.},
  volume = {80},
  issue = {1},
  pages = {141--144},
  numpages = {0},
  year = {1998},
  month = {Jan},
  publisher = {American Physical Society},
  doi = {10.1103/PhysRevLett.80.141},
  url = {https://link.aps.org/doi/10.1103/PhysRevLett.80.141}
}

@article{Ashoori,
  title = {Imaging of low-compressibility strips in the quantum {Hall} liquid},
  author = {Finkelstein, G. and Glicofridis, P.I. and Tessmer, S.H. and Ashoori, R.C. and Melloch, M. R.},
  journal = {Phys. Rev. B},
  volume = {61},
  issue = {24},
  pages = {R16323--R16326},
  numpages = {0},
  year = {2000},
  month = {Jun},
  publisher = {American Physical Society},
  doi = {10.1103/PhysRevB.61.R16323},
  url = {https://link.aps.org/doi/10.1103/PhysRevB.61.R16323}
}

@article{Shytov2,
  title = {Effective action of a compressible quantum {Hall} state edge: Application to tunneling},
  author = {Levitov, L. S. and Shytov, A. V. and Halperin, B. I.},
  journal = {Phys. Rev. B},
  volume = {64},
  issue = {7},
  pages = {075322},
  numpages = {22},
  year = {2001},
  month = {Jul},
  publisher = {American Physical Society},
  doi = {10.1103/PhysRevB.64.075322},
  url = {https://link.aps.org/doi/10.1103/PhysRevB.64.075322}
}

@article{IdrisovPhonon,
  title = {Finite frequency noise in a chiral {Luttinger} liquid coupled to phonons},
  author = {Idrisov, Edvin G.},
  journal = {Phys. Rev. B},
  volume = {100},
  issue = {15},
  pages = {155422},
  numpages = {11},
  year = {2019},
  month = {Oct},
  publisher = {American Physical Society},
  doi = {10.1103/PhysRevB.100.155422},
  url = {https://link.aps.org/doi/10.1103/PhysRevB.100.155422}
}

@article{Mora1,
  title = {Fractionalization and anyonic statistics in the integer quantum {Hall} collider},
  author = {Morel, Tom and Lee, June-Young M. and Sim, H.-S. and Mora, Christophe},
  journal = {Phys. Rev. B},
  volume = {105},
  issue = {7},
  pages = {075433},
  numpages = {14},
  year = {2022},
  month = {Feb},
  publisher = {American Physical Society},
  doi = {10.1103/PhysRevB.105.075433},
  url = {https://link.aps.org/doi/10.1103/PhysRevB.105.075433}
}

@misc{Mora2,
      title={Anyonic exchange in a beam splitter}, 
      author={Christophe Mora},
      year={2022},
      eprint={2212.05123},
      archivePrefix={arXiv}
}

@article{Sim,
  title = {Negative Excess Shot Noise by Anyon Braiding},
  author = {Lee, Byeongmok and Han, Cheolhee and Sim, H.-S.},
  journal = {Phys. Rev. Lett.},
  volume = {123},
  issue = {1},
  pages = {016803},
  numpages = {6},
  year = {2019},
  month = {Jul},
  publisher = {American Physical Society},
  doi = {10.1103/PhysRevLett.123.016803},
  url = {https://link.aps.org/doi/10.1103/PhysRevLett.123.016803}
}

@article{Han,
  title = {Fractional Mutual Statistics on Integer Quantum {Hall} Edges},
  author = {Lee, June-Young M. and Han, Cheolhee and Sim, H.-S.},
  journal = {Phys. Rev. Lett.},
  volume = {125},
  issue = {19},
  pages = {196802},
  numpages = {6},
  year = {2020},
  month = {Nov},
  publisher = {American Physical Society},
  doi = {10.1103/PhysRevLett.125.196802},
  url = {https://link.aps.org/doi/10.1103/PhysRevLett.125.196802}
}

@Article{Lee2022,
author={Lee, June-Young M.
and Sim, H.-S.},
title={Non-Abelian anyon collider},
journal={Nat. Comm.},
year={2022},
month={Nov},
day={04},
volume={13},
number={1},
pages={6660},
issn={2041-1723},
doi={10.1038/s41467-022-34329-y},
url={https://doi.org/10.1038/s41467-022-34329-y}
}

@Article{Lee2023,
author={Lee, June-Young M.
and Hong, Changki
and Alkalay, Tomer
and Schiller, Noam
and Umansky, Vladimir
and Heiblum, Moty
and Oreg, Yuval
and Sim, H.-S.},
title={Partitioning of diluted anyons reveals their braiding statistics},
journal={Nature},
year={2023},
month={May},
day={01},
volume={617},
number={7960},
pages={277-281},
issn={1476-4687},
doi={10.1038/s41586-023-05883-2},
url={https://doi.org/10.1038/s41586-023-05883-2}
}

@article{Feve2023,
  title = {Comparing Fractional Quantum {Hall Laughlin} and {Jain} Topological Orders with the Anyon Collider},
  author = {Ruelle, M. and Frigerio, E. and Berroir, J.-M. and Pla\ifmmode \mbox{\c{c}}\else \c{c}\fi{}ais, B. and Rech, J. and Cavanna, A. and Gennser, U. and Jin, Y. and F\`eve, G.},
  journal = {Phys. Rev. X},
  volume = {13},
  issue = {1},
  pages = {011031},
  numpages = {18},
  year = {2023},
  month = {Mar},
  publisher = {American Physical Society},
  doi = {10.1103/PhysRevX.13.011031},
  url = {https://link.aps.org/doi/10.1103/PhysRevX.13.011031}
}

@inbook{Bartolomei2022,
  author  = {Bartolomei, H. and Kumar, M. and Ruelle, M. and F\`{e}ve, G.},
  title   = {Anyonic and Fermionic Statistics in a Mesoscopic Collider},
  booktitle = {Frank Wilczek: 50 Years of Theoretical Physics},
  editor  = {Niemi, Antti J. and Phua, K. K. and Shapere, Alfred},
  publisher = {World Scientific},
  address = {Singapore},
  year    = {2022},
  month   = mar,
  pages   = {11--36},
  doi     = {10.1142/9789811251948_0003},
  url     = {https://www.worldscientific.com/doi/abs/10.1142/9789811251948_0003},
  isbn    = {978-981-125-517-5}
}

@Article{Nakamura2020,
author={Nakamura, J.
and Liang, S.
and Gardner, G. C.
and Manfra, M. J.},
title={Direct observation of anyonic braiding statistics},
journal={Nat. Phys.},
year={2020},
month={Sep},
day={01},
volume={16},
number={9},
pages={931-936},
issn={1745-2481},
doi={10.1038/s41567-020-1019-1},
url={https://doi.org/10.1038/s41567-020-1019-1}
}

@article{Liang,
  title = {Half-Integer Conductance Plateau at the $\ensuremath{\nu}=2/3$ Fractional Quantum {Hall} State in a Quantum Point Contact},
  author = {Nakamura, J. and Liang, S. and Gardner, G. C. and Manfra, M. J.},
  journal = {Phys. Rev. Lett.},
  volume = {130},
  issue = {7},
  pages = {076205},
  numpages = {7},
  year = {2023},
  month = {Feb},
  publisher = {American Physical Society},
  doi = {10.1103/PhysRevLett.130.076205},
  url = {https://link.aps.org/doi/10.1103/PhysRevLett.130.076205}
}

@article{nakamura2023fabryperot,
  title = {{Fabry-P\'erot} Interferometry at the $\ensuremath{\nu}=2/5$ Fractional Quantum {Hall} State},
  author = {Nakamura, J. and Liang, S. and Gardner, G. C. and Manfra, M. J.},
  journal = {Phys. Rev. X},
  volume = {13},
  issue = {4},
  pages = {041012},
  numpages = {11},
  year = {2023},
  month = {Oct},
  publisher = {American Physical Society},
  doi = {10.1103/PhysRevX.13.041012},
  url = {https://link.aps.org/doi/10.1103/PhysRevX.13.041012}
}

@Article{Glidic2023,
author={Glidic, P.
and Maillet, O.
and Piquard, C.
and Aassime, A.
and Cavanna, A.
and Jin, Y.
and Gennser, U.
and Anthore, A.
and Pierre, F.},
title={Quasiparticle {Andreev} scattering in the $\nu${\thinspace}={\thinspace}1/3 fractional quantum {Hall} regime},
journal={Nat. Comm.},
year={2023},
month={Jan},
day={31},
volume={14},
number={1},
pages={514},
issn={2041-1723},
doi={10.1038/s41467-023-36080-4},
url={https://doi.org/10.1038/s41467-023-36080-4}
}

@article{Glidic,
  title = {Cross-Correlation Investigation of Anyon Statistics in the $\ensuremath{\nu}=1/3$ and $2/5$ Fractional Quantum {Hall} States},
  author = {Glidic, P. and Maillet, O. and Aassime, A. and Piquard, C. and Cavanna, A. and Gennser, U. and Jin, Y. and Anthore, A. and Pierre, F.},
  journal = {Phys. Rev. X},
  volume = {13},
  issue = {1},
  pages = {011030},
  numpages = {19},
  year = {2023},
  month = {Mar},
  publisher = {American Physical Society},
  doi = {10.1103/PhysRevX.13.011030},
  url = {https://link.aps.org/doi/10.1103/PhysRevX.13.011030}
}

@article{Mitali,
  title = {Melting of Interference in the Fractional Quantum {Hall} Effect: Appearance of Neutral Modes},
  author = {Bhattacharyya, Rajarshi and Banerjee, Mitali and Heiblum, Moty and Mahalu, Diana and Umansky, Vladimir},
  journal = {Phys. Rev. Lett.},
  volume = {122},
  issue = {24},
  pages = {246801},
  numpages = {5},
  year = {2019},
  month = {Jun},
  publisher = {American Physical Society},
  doi = {10.1103/PhysRevLett.122.246801},
  url = {https://link.aps.org/doi/10.1103/PhysRevLett.122.246801}
}

@Article{Kundu2023,
author={Kundu, Hemanta Kumar
and Biswas, Sourav
and Ofek, Nissim
and Umansky, Vladimir
and Heiblum, Moty},
title={Anyonic interference and braiding phase in a {Mach-Zehnder} interferometer},
journal={Nat. Phys.},
year={2023},
month={Apr},
day={01},
volume={19},
number={4},
pages={515-521},
issn={1745-2481},
doi={10.1038/s41567-022-01899-z},
url={https://doi.org/10.1038/s41567-022-01899-z}
}

@article{Mitali1,
  author  = {Banerjee, Mitali and Heiblum, Moty and Rosenblatt, Amir and Oreg, Yuval and Feldman, Dima E. and Stern, Ady and Umansky, Vladimir},
  title   = {Observed quantization of anyonic heat flow},
  journal = {Nature},
  volume  = {545},
  number  = {7652},
  pages   = {75--79},
  year    = {2017},
  month   = may,
  doi     = {10.1038/nature22052},
  url     = {https://doi.org/10.1038/nature22052}
}

@article{IdrisovAmmeter,
  author = {Idrisov, Edvin G. and Levkivskyi, Ivan P. and Sukhorukov, Eugene V.},
  title={Quantum ammeter: Measuring full counting statistics of electron currents at quantum timescales},
  journal = {Phys. Rev. B},
  volume = {101},
  issue = {24},
  pages = {245426},
  numpages = {7},
  year = {2020},
  month = {Jun},
  publisher = {American Physical Society},
  doi = {10.1103/PhysRevB.101.245426},
  url = {https://link.aps.org/doi/10.1103/PhysRevB.101.245426}
}

@article{Chamon2,
  author = {de C. Chamon, C. and Freed, D. E. and Wen, X. G.},
  title={Nonequilibrium quantum noise in chiral {Luttinger} liquids},
  journal = {Phys. Rev. B},
  volume = {53},
  issue = {7},
  pages = {4033--4053},
  numpages = {0},
  year = {1996},
  month = {Feb},
  publisher = {American Physical Society},
  doi = {10.1103/PhysRevB.53.4033},
  url = {https://link.aps.org/doi/10.1103/PhysRevB.53.4033}
}

@article {Gwendal2020,
	author = {Bartolomei, H. and Kumar, M. and Bisognin, R. and Marguerite, A. and Berroir, J.-M. and Bocquillon, E. and Pla{\c c}ais, B. and Cavanna, A. and Dong, Q. and Gennser, U. and Jin, Y. and F{\`e}ve, G.},
    title={Fractional statistics in anyon collisions},
	volume = {368},
	number = {6487},
	pages = {173--177},
	year = {2020},
	doi = {10.1126/science.aaz5601},
	publisher = {American Association for the Advancement of Science},
	issn = {0036-8075},
	URL = {https://science.sciencemag.org/content/368/6487/173},
	journal = {Science}
}

@article{Rosenow,
  author = {Rosenow, Bernd and Levkivskyi, Ivan P. and Halperin, Bertrand I.},
  title={Current correlations from a mesoscopic anyon collider},
  journal = {Phys. Rev. Lett.},
  volume = {116},
  issue = {15},
  pages = {156802},
  numpages = {5},
  year = {2016},
  month = {Apr},
  publisher = {American Physical Society},
  doi = {10.1103/PhysRevLett.116.156802},
  url = {https://link.aps.org/doi/10.1103/PhysRevLett.116.156802}
}

@article{GutmanZero,
	doi = {10.1209/0295-5075/90/37003},
	url = {https://doi.org/10.1209/0295-5075/90/37003},
	year = 2010,
	month = {may},
	publisher = {{IOP} Publishing},
	volume = {90},
	number = {3},
	pages = {37003},
	author = {D. B. Gutman and Yuval Gefen and A. D. Mirlin},
    title={Bosonization out of equilibrium},
	journal = {EPL}
}

@article{Gutman,
  author = {Gutman, D. B. and Gefen, Yuval and Mirlin, A. D.},
  title={Bosonization of one-dimensional fermions out of equilibrium},
  journal = {Phys. Rev. B},
  volume = {81},
  issue = {8},
  pages = {085436},
  numpages = {22},
  year = {2010},
  month = {Feb},
  publisher = {American Physical Society},
  doi = {10.1103/PhysRevB.81.085436},
  url = {https://link.aps.org/doi/10.1103/PhysRevB.81.085436}
}

@article{Halperin2002,
  title = {Nonuniversal Behavior of Scattering between Fractional Quantum {Hall} Edges},
  author = {Rosenow, Bernd and Halperin, Bertrand I.},
  journal = {Phys. Rev. Lett.},
  volume = {88},
  issue = {9},
  pages = {096404},
  numpages = {4},
  year = {2002},
  month = {Feb},
  publisher = {American Physical Society},
  doi = {10.1103/PhysRevLett.88.096404},
  url = {https://link.aps.org/doi/10.1103/PhysRevLett.88.096404}
}

@article{Das2012,
  author    = {Anindya Das and Yuval Ronen and Yonatan Most and Yuval Oreg and Moty Heiblum and Hadas Shtrikman},
  title     = {Zero-bias peaks and splitting in an {Al--InAs} nanowire topological superconductor as a signature of {Majorana} fermions},
  journal   = {Nat. Phys.},
  year      = {2012},
  volume    = {8},
  pages     = {887--895},
  doi       = {10.1038/nphys2479},
  url       = {https://doi.org/10.1038/nphys2479}
}

@article{Dolev2008,
  author    = {Marcin Dolev and Moty Heiblum and Vladimir Umansky and Ady Stern and Diana Mahalu},
  title     = {Observation of a quarter of an electron charge at the $\nu = 5/2$ quantum {Hall} state},
  journal   = {Nature},
  year      = {2008},
  volume    = {452},
  pages     = {829--834},
  doi       = {10.1038/nature06855},
  url       = {https://doi.org/10.1038/nature06855}
}

@article{Banerjee2018,
  author  = {Banerjee, Mitali and Heiblum, Moty and Umansky, Vladimir and Feldman, Dima E. and Oreg, Yuval and Stern, Ady},
  title   = {Observation of half-integer thermal {Hall} conductance},
  journal = {Nature},
  volume  = {559},
  number  = {7713},
  pages   = {205--210},
  year    = {2018},
  doi     = {10.1038/s41586-018-0184-1},
}

@article{Kane_2003,
  title = {Shot noise and the transmission of dilute {Laughlin} quasiparticles},
  author = {Kane, C. L. and Fisher, Matthew P. A.},
  journal = {Phys. Rev. B},
  volume = {67},
  issue = {4},
  pages = {045307},
  numpages = {17},
  year = {2003},
  month = {Jan},
  publisher = {American Physical Society},
  doi = {10.1103/PhysRevB.67.045307},
  url = {https://link.aps.org/doi/10.1103/PhysRevB.67.045307}
}

@article{Ivan_Noneq_Bosonization,
  title = {Universal nonequilibrium states at the fractional quantum {Hall} edge},
  author = {Levkivskyi, Ivan P.},
  journal = {Phys. Rev. B},
  volume = {93},
  issue = {16},
  pages = {165427},
  numpages = {6},
  year = {2016},
  month = {Apr},
  publisher = {American Physical Society},
  doi = {10.1103/PhysRevB.93.165427},
  url = {https://link.aps.org/doi/10.1103/PhysRevB.93.165427}
}

@article{Spanton2017Observation,title={Observation of fractional {Chern} insulators in a {van der Waals} heterostructure},author={E. Spanton and A. Zibrov and Haoxin Zhou and T. Taniguchi and Kenji Watanabe and M. Zaletel and A. Young},journal={Science},year={2018},volume={360},pages={62 - 66},doi={10.1126/science.aan8458}}

@article{Peterson2021Trapped,title={Trapped fractional charges at bulk defects in topological insulators},author={Christopher W. Peterson and Tianhe Li and Wentao Jiang and T. Hughes and G. Bahl},journal={Nature},year={2021},volume={589},pages={376 - 380},doi={10.1038/s41586-020-03117-3}}

@article{Zhang2020Experimental,title={Experimental Observation of Higher-Order Topological {Anderson} Insulators},author={Weixuan Zhang and Deyuan Zou and Qingsong Pei and Wenjing He and Jiacheng Bao and Houjun Sun and Xiangdong Zhang},journal={Phys. Rev. Lett.},year={2021},volume={126},pages={ 146802 },doi={10.1103/physrevlett.126.146802}}

@article{Breunig2021Opportunities,
title={Opportunities in topological insulator devices},
author={O. Breunig and Y. Ando},
journal={Nat. Rev. Phys.},
year={2021},
volume={4},
pages={184 - 193},
doi={10.1038/s42254-021-00402-6}
}

@article{Xu2023Observation,
  title = {Observation of integer and fractional quantum anomalous {Hall} effects in Twisted Bilayer {$\ensuremath{\text{MoTe}_2}$}},
  author = {Xu, Fan and Sun, Zheng and Jia, Tongtong and Liu, Chang and Xu, Cheng and Li, Chushan and Gu, Yu and Watanabe, Kenji and Taniguchi, Takashi and Tong, Bingbing and Jia, Jinfeng and Shi, Zhiwen and Jiang, Shengwei and Zhang, Yang and Liu, Xiaoxue and Li, Tingxin},
  year = 2023,
  month = sep,
  journal = {Phys. Rev. X},
  volume = {13},
  number = {3},
  pages = {031037},
  publisher = {American Physical Society},
  doi = {10.1103/PhysRevX.13.031037},
  url = {https://link.aps.org/doi/10.1103/PhysRevX.13.031037},
  urldate = {2026-02-27}
}

@article{Nagaosa2009Anomalous,title={Anomalous {Hall} effect},author={N. Nagaosa and J. Sinova and S. Onoda and A. {MacDonald} and N. Ong},journal={Rev. Mod. Phys.},year={2010},volume={82},pages={1539-1592},doi={10.1103/revmodphys.82.1539}}

@article{Carrega2021Anyons,
title={Anyons in quantum {Hall} interferometry},
author={M. Carrega and L. Chirolli and S. Heun and L. Sorba},
journal={Nat. Rev. Phys.},
year={2021},
volume={3},
pages={698 - 711},
doi={10.1038/s42254-021-00351-0}
}

@article{Schiller2022Anyon,title={Anyon Statistics through Conductance Measurements of Time-Domain Interferometry},author={N. Schiller and Yotam Shapira and A. Stern and Y. Oreg},journal={Phys. Rev. Lett.},year={2023},volume={131},pages={ 186601 },doi={10.1103/physrevlett.131.186601}}

@article{Saminadayar1997,
  title = {Observation of the $e/3$ Fractionally Charged {Laughlin} Quasiparticle},
  author = {Saminadayar, L. and Glattli, D. C. and Jin, Y. and Etienne, B.},
  journal = {Phys. Rev. Lett.},
  volume = {79},
  number = {13},
  pages = {2526--2529},
  year = {1997},
  doi = {10.1103/PhysRevLett.79.2526}
}

@article{dePicciotto1997,
  title = {Direct observation of a fractional charge},
  author = {de-Picciotto, R. and Reznikov, M. and Heiblum, M. and Umansky, V. and Bunin, G. and Mahalu, D.},
  journal = {Nature},
  volume = {389},
  pages = {162--164},
  year = {1997},
  doi = {10.1038/38241}
}

@article{Biswas2022,
  title = {Shot noise does not always provide the quasiparticle charge},
  author = {Biswas, S. and Bhattacharyya, R. and Kundu, H. K. and Das, A. and Heiblum, M. and Umansky, V. and Goldstein, M. and Gefen, Y.},
  journal = {Nat. Phys.},
  volume = {18},
  pages = {1476--1482},
  year = {2022},
  doi = {10.1038/s41567-022-01666-5}
}

@article{Goldman1995,
  title = {Resonant tunneling in the quantum {Hall} regime: Measurement of fractional charge},
  author = {Goldman, V. J. and Su, B.},
  journal = {Science},
  volume = {267},
  number = {5200},
  pages = {1010--1012},
  year = {1995},
  doi = {10.1126/science.267.5200.1010}
}

@article{Ponomarenko2024Unusual,title={Unusual Quasiparticles and Tunneling Conductance in Quantum Point Contacts in $\nu=2/3$ Fractional Quantum {Hall} Systems},author={Vadim Ponomarenko and Yuli Lyanda-Geller},journal={Phys. Rev. Lett.},year={2024},volume={133 },pages={ 076503 },doi={10.1103/physrevlett.133.076503}}

@article{Kiczynski2022Engineering,title={Engineering topological states in atom-based semiconductor quantum dots},author={M. Kiczynski and S. K. Gorman and H. Geng and M. Donnelly and Y. Chung and Y. He and J. Keizer and M. Simmons},journal={Nature},year={2022},volume={606},pages={694 - 699},doi={10.1038/s41586-022-04706-0}}

@article{Miller2007Fractional,title={Fractional quantum {Hall} effect in a quantum point contact at filling fraction 5/2},author={Jeffrey Boone Miller and I. Radu and D. Zumbühl  and E. Levenson-Falk and M. Kastner and C. Marcus and L. Pfeiffer and K. West},journal={Nat. Phys.},year={2007},volume={3},pages={561-565},doi={10.1038/nphys658}}

@article{PhysRevLett.132.216601,
  title = {Finite Width of Anyons Changes Their Braiding Signature},
  author = {Iyer, K. and Ronetti, F. and Gr\'emaud, B. and Martin, T. and Rech, J. and Jonckheere, T.},
  journal = {Phys. Rev. Lett.},
  volume = {132},
  issue = {21},
  pages = {216601},
  numpages = {6},
  year = {2024},
  month = {May},
  publisher = {American Physical Society},
  doi = {10.1103/PhysRevLett.132.216601},
  url = {https://link.aps.org/doi/10.1103/PhysRevLett.132.216601}
}

@article{Han2016,
  author = {Han, Cheolhee and Park, Jinhong and Gefen, Yuval and Sim, H.-S.},
  title = {Topological vacuum bubbles by anyon braiding},
  journal = {Nat. Comm.},
  volume = {7},
  number = {1},
  pages = {11131},
  year = {2016},
  month = jun,
  day = {02},
  doi = {10.1038/ncomms11131},
  url = {https://doi.org/10.1038/ncomms11131},
  publisher = {Nature Publishing Group}
}

@misc{Rosenow2025,
  author = {Bernd Rosenow and Bertrand I. Halperin},
  title = {Braids and Beams: Exploring Fractional Statistics with Mesoscopic Anyon Colliders},
  year = {2025},
  eprint = {2510.04319},
  archivePrefix = {arXiv}
}

@article{two_point_contact_interferometer,
  title = {Two point-contact interferometer for quantum {Hall} systems},
  author = {de C. Chamon, C. and Freed, D. E. and Kivelson, S. A. and Sondhi, S. L. and Wen, X. G.},
  journal = {Phys. Rev. B},
  volume = {55},
  issue = {4},
  pages = {2331--2343},
  numpages = {0},
  year = {1997},
  month = {Jan},
  publisher = {American Physical Society},
  doi = {10.1103/PhysRevB.55.2331},
  url = {https://link.aps.org/doi/10.1103/PhysRevB.55.2331}
}

@article{Kambly2010Factorial,
  title = {Factorial cumulants reveal interactions in counting statistics},
  author = {Kambly, Dania and Flindt, Christian and B\"uttiker, Markus},
  journal = {Phys. Rev. B},
  volume = {83},
  issue = {7},
  pages = {075432},
  numpages = {11},
  year = {2011},
  month = {Feb},
  publisher = {American Physical Society},
  doi = {10.1103/PhysRevB.83.075432},
  url = {https://link.aps.org/doi/10.1103/PhysRevB.83.075432}
}

@article{Landi2023Current,
  title = {Current {{Fluctuations}} in {{Open Quantum Systems}}: {{Bridging}} the {{Gap Between Quantum Continuous Measurements}} and {{Full Counting Statistics}}},
  shorttitle = {Current {{Fluctuations}} in {{Open Quantum Systems}}},
  author = {Landi, Gabriel T. and Kewming, Michael J. and Mitchison, Mark T. and Potts, Patrick P.},
  year = 2024,
  month = apr,
  journal = {PRX Quantum},
  volume = {5},
  number = {2},
  pages = {020201},
  publisher = {American Physical Society},
  doi = {10.1103/PRXQuantum.5.020201},
  url = {https://link.aps.org/doi/10.1103/PRXQuantum.5.020201},
  urldate = {2026-02-27}
}

@article{PhysRevLett.105.256802,
  title = {Full Counting Statistics of a {Luttinger} Liquid Conductor},
  author = {Gutman, D. B. and Gefen, Yuval and Mirlin, A. D.},
  journal = {Phys. Rev. Lett.},
  volume = {105},
  issue = {25},
  pages = {256802},
  numpages = {4},
  year = {2010},
  month = {Dec},
  publisher = {American Physical Society},
  doi = {10.1103/PhysRevLett.105.256802},
  url = {https://link.aps.org/doi/10.1103/PhysRevLett.105.256802}
}

@article{PhysRevB.87.195433,
  title = {Analytically solvable model of an electronic {Mach-Zehnder} interferometer},
  author = {Ngo Dinh, St\'ephane and Bagrets, Dmitry A. and Mirlin, Alexander D.},
  journal = {Phys. Rev. B},
  volume = {87},
  issue = {19},
  pages = {195433},
  numpages = {26},
  year = {2013},
  month = {May},
  publisher = {American Physical Society},
  doi = {10.1103/PhysRevB.87.195433},
  url = {https://link.aps.org/doi/10.1103/PhysRevB.87.195433}
}

@article{PhysRevB.91.245419,
  title = {Counting statistics and dephasing transition in an electronic {Mach-Zehnder} interferometer},
  author = {Helzel, A. and Litvin, L. V. and Levkivskyi, I. P. and Sukhorukov, E. V. and Wegscheider, W. and Strunk, C.},
  journal = {Phys. Rev. B},
  volume = {91},
  issue = {24},
  pages = {245419},
  numpages = {13},
  year = {2015},
  month = {Jun},
  publisher = {American Physical Society},
  doi = {10.1103/PhysRevB.91.245419},
  url = {https://link.aps.org/doi/10.1103/PhysRevB.91.245419}
}

@book{Giamarchi2004,
  author = {Giamarchi, Thierry},
  title = {Quantum Physics in One Dimension},
  publisher = {Clarendon Press},
  address = {Oxford},
  year = {2004},
  isbn = {978-0-19-852500-4}
}

@article{Bisognin2019,
  author = {Bisognin, R. and Bartolomei, H. and Kumar, M. and Safi, I. and Berroir, J.-M. and Bocquillon, E. and Degiovanni, P. and Cavanna, A. and Jin, Y. and F{\`e}ve, G.},
  title = {Microwave photons emitted by fractionally charged quasiparticles},
  journal = {Nat. Comm.},
  volume = {10},
  number = {1},
  pages = {1708},
  year = {2019},
  month = apr,
  day = {12},
  doi = {10.1038/s41467-019-09758-x},
  url = {https://doi.org/10.1038/s41467-019-09758-x},
  publisher = {Springer Nature}
}

@article{Leinaas1977,
  author = {Leinaas, J. M. and Myrheim, J.},
  title = {On the theory of identical particles},
  journal = {Nuovo Cimento B},
  volume = {37},
  pages = {1--23},
  year = {1977},
  doi = {10.1007/BF02727953}
}

@article{Wilczek1982,
  author = {Wilczek, Frank},
  title = {Quantum mechanics of fractional-spin particles},
  journal = {Phys. Rev. Lett.},
  volume = {49},
  pages = {957--959},
  year = {1982},
  doi = {10.1103/PhysRevLett.49.957}
}

@article{Wilczek1982Anyon,
  author = {Wilczek, Frank},
  title = {Magnetic flux, angular momentum, and statistics},
  journal = {Phys. Rev. Lett.},
  volume = {48},
  pages = {1144--1146},
  year = {1982},
  doi = {10.1103/PhysRevLett.48.1144}
}

@article{Maciejko2010,
  title = {Fractional Topological Insulators in Three Dimensions},
  author = {Maciejko, Joseph and Qi, Xiao-Liang and Karch, Andreas and Zhang, Shou-Cheng},
  journal = {Phys. Rev. Lett.},
  volume = {105},
  issue = {24},
  pages = {246809},
  numpages = {4},
  year = {2010},
  month = {Dec},
  publisher = {American Physical Society},
  doi = {10.1103/PhysRevLett.105.246809},
  url = {https://link.aps.org/doi/10.1103/PhysRevLett.105.246809}
}

@article{Kane1992,
  author = {Kane, C. L. and Fisher, Matthew P. A.},
  title = {Transmission through barriers and resonant tunneling in an interacting one-dimensional electron gas},
  journal = {Phys. Rev. B},
  volume = {46},
  pages = {15233--15243},
  year = {1992},
  doi = {10.1103/PhysRevB.46.15233}
}

@article{Werkmeister2025,
  author  = {Werkmeister, Thomas and Ehrets, James R. and Wesson, Marie E. and Najafabadi, Danial H. and Watanabe, Kenji and Taniguchi, Takashi and Halperin, Bertrand I. and Yacoby, Amir and Kim, Philip},
  title   = {Anyon braiding and telegraph noise in a graphene interferometer},
  journal = {Science},
  volume  = {388},
  number  = {6748},
  pages   = {730--735},
  year    = {2025},
  doi     = {10.1126/science.adp5015},
}

@article{MZI_interferometer,
  title = {{Mach-Zehnder} interferometry of fractional quantum {Hall} edge states},
  author = {Levkivskyi, Ivan P. and Boyarsky, Alexey and Fr\"ohlich, J\"urg and Sukhorukov, Eugene V.},
  journal = {Phys. Rev. B},
  volume = {80},
  issue = {4},
  pages = {045319},
  numpages = {21},
  year = {2009},
  month = {Jul},
  publisher = {American Physical Society},
  doi = {10.1103/PhysRevB.80.045319},
  url = {https://link.aps.org/doi/10.1103/PhysRevB.80.045319}
}

@article{PhysRevB.86.245105,
  title = {Theory of fractional quantum {Hall} interferometers},
  author = {Levkivskyi, Ivan P. and Fr\"ohlich, J\"urg and Sukhorukov, Eugene V.},
  journal = {Phys. Rev. B},
  volume = {86},
  issue = {24},
  pages = {245105},
  numpages = {21},
  year = {2012},
  month = {Dec},
  publisher = {American Physical Society},
  doi = {10.1103/PhysRevB.86.245105},
  url = {https://link.aps.org/doi/10.1103/PhysRevB.86.245105}
}

@article{noise_induced_PT,
  title = {Noise-Induced Phase Transition in the Electronic {Mach-Zehnder} Interferometer},
  author = {Levkivskyi, Ivan P. and Sukhorukov, Eugene V.},
  journal = {Phys. Rev. Lett.},
  volume = {103},
  issue = {3},
  pages = {036801},
  numpages = {4},
  year = {2009},
  month = {Jul},
  publisher = {American Physical Society},
  doi = {10.1103/PhysRevLett.103.036801},
  url = {https://link.aps.org/doi/10.1103/PhysRevLett.103.036801}
}

@misc{zhang2025effective,
  author = {Zhang, Gu and Gornyi, Igor and Gefen, Yuval},
  title = {Effective linear response in non-equilibrium anyonic systems},
  year = {2025},
  eprint={2510.03985},
  archivePrefix = {arXiv}
}

@article{PhysRevLett.134.096303,
  title = {Landscapes of an Out-of-Equilibrium Anyonic Sea},
  author = {Zhang, Gu and Gornyi, Igor and Gefen, Yuval},
  journal = {Phys. Rev. Lett.},
  volume = {134},
  issue = {9},
  pages = {096303},
  numpages = {8},
  year = {2025},
  month = {Mar},
  publisher = {American Physical Society},
  doi = {10.1103/PhysRevLett.134.096303},
  url = {https://link.aps.org/doi/10.1103/PhysRevLett.134.096303}
}

@article{PhysRevB.83.155440,
  title = {Theory of the {Fabry-P\'erot} quantum {Hall} interferometer},
  author = {Halperin, Bertrand I. and Stern, Ady and Neder, Izhar and Rosenow, Bernd},
  journal = {Phys. Rev. B},
  volume = {83},
  issue = {15},
  pages = {155440},
  numpages = {17},
  year = {2011},
  month = {Apr},
  publisher = {American Physical Society},
  doi = {10.1103/PhysRevB.83.155440},
  url = {https://link.aps.org/doi/10.1103/PhysRevB.83.155440}
}

@article{PhysRevLett.98.106801,
  title = {Influence of Interactions on Flux and Back-Gate Period of Quantum {Hall} Interferometers},
  author = {Rosenow, B. and Halperin, B. I.},
  journal = {Phys. Rev. Lett.},
  volume = {98},
  issue = {10},
  pages = {106801},
  numpages = {4},
  year = {2007},
  month = {Mar},
  publisher = {American Physical Society},
  doi = {10.1103/PhysRevLett.98.106801},
  url = {https://link.aps.org/doi/10.1103/PhysRevLett.98.106801}
}

@article{feldman_robustness_2022,
    title = {Robustness of quantum {Hall} interferometry},
    volume = {105},
    issn = {2469-9950, 2469-9969},
    url = {https://link.aps.org/doi/10.1103/PhysRevB.105.165310},
    doi = {10.1103/PhysRevB.105.165310},
    number = {16},
    urldate = {2025-12-23},
    journal = {Phys. Rev. B},
    author = {Feldman, D. E. and Halperin, Bertrand I.},
    month = apr,
    year = {2022},
    pages = {165310},
}

@book{Ezawa2013,
  author    = {Ezawa, Zyun Francis},
  title     = {Quantum {Hall} Effects: Recent Theoretical and Experimental Developments},
  edition   = {3},
  publisher = {World Scientific},
  year      = {2013},
  month     = may,
  doi       = {10.1142/8210},
  isbn      = {978-981-4360-75-3},
  url       = {https://www.worldscientific.com/worldscibooks/10.1142/8210}
}

@article{PhysRevB.110.L041402,
  title = {Quantization of the anyonic spectral density of heat currents},
  author = {Idrisov, Edvin G. and Ekstr\"om, Johan},
  journal = {Phys. Rev. B},
  volume = {110},
  issue = {4},
  pages = {L041402},
  numpages = {4},
  year = {2024},
  month = {Jul},
  publisher = {American Physical Society},
  doi = {10.1103/PhysRevB.110.L041402},
  url = {https://link.aps.org/doi/10.1103/PhysRevB.110.L041402}
}

@misc{DLMF,
  author={{NIST Digital Library of Mathematical Functions}},
  howpublished = {\url{https://dlmf.nist.gov/}, Release 1.2.7},
  note         = {{F}.~W.~J. Olver, A.~B. {Olde Daalhuis}, D.~W. Lozier, B.~I. Schneider,
                  R.~F. Boisvert, C.~W. Clark, B.~R. Miller, B.~V. Saunders,
                  H.~S. Cohl, and M.~A. McClain, eds.},
    year={2026}
}

@article{PhysRevLett.130.186203,
  title   = {Anyonic statistics revealed by the {Hong-Ou-Mandel} dip for fractional excitations},
  author  = {Jonckheere, T. and Rech, J. and Gr\'emaud, B. and Martin, T.},
  journal = {Phys. Rev. Lett.},
  volume  = {130},
  pages   = {186203},
  year    = {2023},
  doi     = {10.1103/PhysRevLett.130.186203}
}

@article{Ofek2010,
  title   = {Role of interactions in an electronic {Fabry--Perot} interferometer
             operating in the quantum {Hall} effect regime},
  author  = {Ofek, Nissim and Bid, Aveek and Heiblum, Moty and Stern, Ady
             and Umansky, Vladimir and Mahalu, Diana},
  journal = {Proc. Natl. Acad. Sci. U.S.A.},
  volume  = {107},
  pages   = {5276--5281},
  year    = {2010},
  doi     = {10.1073/pnas.0912624107}
}

@article{McClure2009,
  title   = {Edge-state velocity and coherence in a quantum {Hall}
             {Fabry--P\'erot} interferometer},
  author  = {McClure, D. T. and Zhang, Yiming and Rosenow, B. and
             Levenson-Falk, E. M. and Marcus, C. M. and Pfeiffer, L. N.
             and West, K. W.},
  journal = {Phys. Rev. Lett.},
  volume  = {103},
  pages   = {206806},
  year    = {2009},
  doi     = {10.1103/PhysRevLett.103.206806}
}

@article{Deprez2021,
  title   = {A tunable {Fabry--P\'erot} quantum {Hall} interferometer in graphene},
  author  = {D\'eprez, Corentin and Veyrat, Louis and Vignaud, Hadrien and
             Nayak, Goutham and Watanabe, Kenji and Taniguchi, Takashi and
             Gay, Fr\'ed\'eric and Sellier, Hermann and Sac\'ep\'e, Benjamin},
  journal = {Nat. Nanotechnol.},
  volume  = {16},
  pages   = {555--562},
  year    = {2021},
  doi     = {10.1038/s41565-021-00847-x}
}

@article{Kivelson2025,
  title   = {Modified interferometer to measure anyonic braiding statistics},
  author  = {Kivelson, S. A. and Murthy, C.},
  journal = {Phys. Rev. Lett.},
  volume  = {135},
  pages   = {126605},
  year    = {2025},
  doi     = {10.1103/x4w5-h3bb}
}

@article{Law2006,
  title   = {Electronic {Mach-Zehnder} interferometer as a tool to probe fractional statistics},
  author  = {Law, K. T. and Feldman, D. E. and Gefen, Yuval},
  journal = {Phys. Rev. B},
  volume  = {74},
  pages   = {045319},
  year    = {2006},
  doi     = {10.1103/PhysRevB.74.045319}
}

@article{Campagnano2012,
  title   = {{Hanbury Brown--Twiss} interference of anyons},
  author  = {Campagnano, Gabriele and Zilberberg, Oded and Gornyi, Igor V. and
             Feldman, D. E. and Potter, Andrew C. and Gefen, Yuval},
  journal = {Phys. Rev. Lett.},
  volume  = {109},
  pages   = {106802},
  year    = {2012},
  doi     = {10.1103/PhysRevLett.109.106802}
}

@article{Ji2003,
  title   = {An electronic {Mach--Zehnder} interferometer},
  author  = {Ji, Yang and Chung, Yunchul and Sprinzak, D. and Heiblum, M. and
             Mahalu, D. and Shtrikman, Hadas},
  journal = {Nature},
  volume  = {422},
  pages   = {415--418},
  year    = {2003},
  doi     = {10.1038/nature01503}
}

@article{Safi2001,
  title   = {Partition noise and statistics in the fractional quantum {Hall} effect},
  author  = {Safi, I. and Devillard, P. and Martin, T.},
  journal = {Phys. Rev. Lett.},
  volume  = {86},
  pages   = {4628--4631},
  year    = {2001},
  doi     = {10.1103/PhysRevLett.86.4628}
}

@article{Samuelson2026,
  title   = {Slow quasiparticle dynamics and anyonic statistics in a fractional
             quantum {Hall} {Fabry-P\'erot} interferometer},
  author  = {Samuelson, N. L. and Cohen, L. A. and Wang, W. and Blanch, S. and
             Taniguchi, T. and Watanabe, K. and Zaletel, M. P. and Young, A. F.},
  journal = {Phys. Rev. X},
  volume  = {16},
  pages   = {011062},
  year    = {2026},
  doi     = {10.1103/fwjg-mx9h}
}

@article{Ghosh2025Bunching,
  author  = {Ghosh, Bikash and Labendik, Maria and Umansky, Vladimir and Heiblum, Moty and Mross, David F.},
  title   = {Coherent bunching of anyons and dissociation in an interference experiment},
  journal = {Nature},
  volume  = {642},
  number  = {8069},
  pages   = {922--927},
  year    = {2025},
  doi     = {10.1038/s41586-025-09143-3},
}

@article{Ghosh2025MZ,
  title   = {Anyonic braiding in a chiral {Mach--Zehnder} interferometer},
  author  = {Ghosh, Bikash and Labendik, Maria and Musina, Liliia and Umansky, Vladimir and Heiblum, Moty and Mross, David F.},
  journal = {Nat. Phys.},
  volume  = {21},
  pages   = {1392--1397},
  year    = {2025},
  doi     = {10.1038/s41567-025-02960-3}
}

@article{Kim2024,
  title   = {{Aharonov--Bohm} interference and statistical phase-jump evolution in fractional quantum {Hall} states in bilayer graphene},
  author  = {Kim, Jehyun and Dev, Himanshu and Kumar, Ravi and Ilin, Alexey and Haug, Andr\'{e} and Bhardwaj, Vishal and Hong, Changki and Watanabe, Kenji and Taniguchi, Takashi and Stern, Ady and Ronen, Yuval},
  journal = {Nature Nanotechnology},
  volume  = {19},
  number  = {11},
  pages   = {1619--1626},
  year    = {2024},
  doi     = {10.1038/s41565-024-01751-w},
  publisher = {Nature Publishing Group}
}

@article{PhysRevLett.132.156501,
  title   = {Effect of the soliton width on nonequilibrium exchange phases of anyons},
  author  = {Thamm, Matthias and Rosenow, Bernd},
  journal = {Phys. Rev. Lett.},
  volume  = {132},
  pages   = {156501},
  year    = {2024},
  doi     = {10.1103/PhysRevLett.132.156501}
}

@article{Feldman2021Fractional,
  title   = {Fractional charge and fractional statistics in the quantum {Hall} effects},
  author  = {Feldman, D. E. and Halperin, Bertrand I.},
  journal = {Rep. Prog. Phys.},
  volume  = {84},
  pages   = {076501},
  year    = {2021},
  doi     = {10.1088/1361-6633/ac03aa}
}

@article{Ruelle2025,
  title   = {Time-domain braiding of anyons},
  author  = {Ruelle, M. and Frigerio, E. and Baudin, E. and Berroir, J.-M. and
             Pla{\c c}ais, B. and Gr\'emaud, B. and Jonckheere, T. and Martin, T. and
             Rech, J. and Cavanna, A. and Gennser, U. and Jin, Y. and
             M\'enard, G. and F\`eve, G.},
  journal = {Science},
  volume  = {389},
  number  = {6755},
  eid     = {7695},
  pages   = {7695},
  year    = {2025},
  doi     = {10.1126/science.adm7695},
}

@article{Taktak2022,
  title   = {Two-particle time-domain interferometry in the fractional quantum {Hall} effect regime},
  author  = {Taktak, I. and Kapfer, M. and Nath, J. and Roulleau, P. and
             Acciai, M. and Splettstoesser, J. and Farrer, I. and
             Ritchie, D. A. and Glattli, D. C.},
  journal = {Nat. Comm.},
  volume  = {13},
  pages   = {5863},
  year    = {2022},
  doi={https://doi.org/10.1038/s41467-022-33603-3}
}

@article{Witten1989,
  author  = {Witten, Edward},
  title   = {Quantum field theory and the {J}ones polynomial},
  journal = {Communications in Mathematical Physics},
  year    = {1989},
  volume  = {121},
  number  = {3},
  pages   = {351--399},
  doi     = {10.1007/BF01217730}
}

@article{Polyakov1988,
  author  = {Polyakov, A. M.},
  title   = {Fermi--{B}ose transmutation induced by gauge fields},
  journal = {Modern Physics Letters A},
  year    = {1988},
  volume  = {3},
  number  = {3},
  pages   = {325--328},
  doi     = {10.1142/S0217732388000398}
}

@article{RevModPhys.80.1083,
  title = {Non-Abelian anyons and topological quantum computation},
  author = {Nayak, Chetan and Simon, Steven H. and Stern, Ady and Freedman, Michael and Das Sarma, Sankar},
  journal = {Rev. Mod. Phys.},
  volume = {80},
  issue = {3},
  pages = {1083--1159},
  numpages = {0},
  year = {2008},
  month = {Sep},
  publisher = {American Physical Society},
  doi = {10.1103/RevModPhys.80.1083},
  url = {https://link.aps.org/doi/10.1103/RevModPhys.80.1083}
}
\end{document}